\documentclass[12pt]{article}
\usepackage[utf8]{inputenc}
\usepackage{amsmath, amsthm, bm, amssymb, hyperref, xcolor, caption, subcaption, graphicx, dsfont, multirow, booktabs, array, rotating}

\hypersetup{
	hidelinks,
    colorlinks = true,
    linkbordercolor = {white},
    citecolor = blue,
    urlcolor = blue,
    linkcolor = blue,
    linktoc = all 
}

\usepackage[
  top=1in, 
  bottom=1in,
  left=1in,
  right=1in
]{geometry} 

\usepackage{natbib}
\setcitestyle{authoryear}
\title{\textsc{When Respondents Don't Care Anymore: Identifying the Onset of Careless Responding}}
\author{Max Welz \\
  Department of Econometrics,
  Erasmus University Rotterdam \\
  Department of Psychology,
  University of Zurich\\
  \href{mailto:max.welz@uzh.ch}{\texttt{max.welz@uzh.ch}}
  \and
  Andreas Alfons \\
  Department of Econometrics,
  Erasmus University Rotterdam \\
  \href{mailto:alfons@ese.eur.nl}{\texttt{alfons@ese.eur.nl}}}
  
\date{}
    
\usepackage{tikz, latexsym}
    \usetikzlibrary{decorations.pathreplacing}
    \usetikzlibrary{fadings}
    
\newcommand{\Blowdim}{\boldsymbol{S}}

\newcommand{\Bf}{\boldsymbol{f}}
\newcommand{\Btheta}{\boldsymbol{\theta}}
\newcommand{\Ell}{\mathcal{L}}

\newcommand{\N}{\mathbb{N}}

\newcommand{\RE}{RE}
\definecolor{EURwarmgray}{RGB}{227,218,216}

\newcommand{\obs}[1]{\mathbf{#1}}
\newcommand{\vect}[1]{\boldsymbol{#1}}
\newcommand{\mat}[1]{\boldsymbol{#1}}

{
	\theoremstyle{definition}
	\newtheorem{assumption}{Assumption}
}

\usepackage[ruled, linesnumbered]{algorithm2e}
\SetKw{KwBy}{by}
\SetKwInput{KwInp}{Initialization}

\SetCommentSty{mycommfont}

\usepackage{grffile}

\newcommand{\pkg}[1]{\texttt{#1}}
\newcommand{\proglang}[1]{\textsf{#1}}
\DeclareMathOperator*{\argmax}{arg\,max}

\usepackage{setspace}
\newcolumntype{L}[1]{>{\raggedright\arraybackslash}m{#1}}

\makeatletter
\let\ftype@table\ftype@figure
\makeatother

\begin{document}

{\singlespacing
\maketitle

\begin{abstract}
\noindent 
Questionnaires in the behavioral sciences tend to be lengthy. However, literature suggests that survey length is a contributing factor to careless responding, with longer questionnaires yielding higher probability that participants start responding carelessly. Consequently, in long surveys a large number of participants may engage in careless responding, posing a major threat to internal validity. We propose a novel method for identifying the onset of careless responding (or an absence thereof) that searches for a changepoint in combined measurements of multiple dimensions in which carelessness may manifest, such as inconsistency and invariability. It is highly flexible, based on machine learning, and provides statistical guarantees for controlling the false positive rate. In simulation experiments, the proposed method achieves high accuracy in identifying carelessness onset and discriminates well between attentive and various types of careless responding, even when a large number of careless respondents are present. An empirical application highlights how identifying partial carelessness uncovers novel insights on careless responding behavior. Furthermore, we provide the freely available open source software package \pkg{carelessonset} to facilitate adoption by empirical researchers.
\end{abstract}

}

\newpage


\section*{Introduction}
Research in the behavioral sciences often involves the administration of lengthy (online) self-report questionnaires. For instance, common personality measures tend to consist of hundreds of items, such as the \textit{Revised NEO Personality Inventory} \citep{costa1992revised} with 240 items or the \textit{Minnesota Multiphasic Personality Inventory-2 Restructured Form} \citep{benporath2008} with 338 items. Even if one does not use such extensive measures, the number of items can easily reach three digits by including several shorter measures. However, it is well known that questionnaire length can have a concerning adverse effect on measurement accuracy: Respondents may experience fatigue or boredom as they progress through a lengthy questionnaire, which can provoke careless responding \citep[e.g.,][]{bowling2021length,gibson2020,galesic2009}, in particular in online administrations \citep[e.g.,][]{johnson2005,meade2012}. We call such participants \textit{partially careless respondents} as they initially provide accurate responses, but resort to careless responding after a certain item and remain careless for the remainder of the questionnaire.

According to the definition of \citet{ward2023}, \emph{``careless responding occurs when participants are not basing their response on the item content, and it can occur when a respondent does not read an item, does not understand an item, or is unmotivated to think about what the item is asking''.}\footnote{Perhaps more commonly used, \citet{huang2012} define careless responding as \textit{``a response set in which the respondent answers a survey measure with low or little motivation to comply with survey instructions, correctly interpret item content, and provide accurate responses''}. We prefer the definition of \citet{ward2023}, as it highlights that careless responding may not be intentional \citep[cf.][]{huang2012, ward2015, ward2017}. Also note that \citet{huang2012} refer to careless responding as \textit{insufficient effort responding} in their definition. Other synonyms for careless responding are \emph{participant inattention} \citep{maniaci2014}, \emph{inconsistent responding} \citep{greene1978}, \emph{protocol invalidity} \citep{johnson2005}, and \emph{random responding} \citep[e.g.,][]{beach1989,berry1992,emons2008}. Notably, \citet{schroeders2022} point out that the latter---random responding---may be a misnomer since careless responding can also arise in non-random patterns (e.g., a recurring sequence of 1-2-3-4-5).}
Careless responding has been identified as a major threat to the validity of research findings through a variety of psychometric and statistical issues such as errors in statistical hypothesis testing, deteriorated model fit, and lower scale reliability \citep[e.g.,][]{stosic2024, voss2024, arias2020,huang2012,huang2015}. It is widely prevalent \citep[e.g.,][]{bowling2016,curran2016,desimone2015,meade2012}, suspected to be present in all questionnaire data \citep{ward2023}, and already incidence rates as low as~5\% can be problematic \citep{arias2020, crede2010, welz2024maxbias, welz2024polycor}. Often viewed as merely an insidious source of measurement error, careless responding can be a valuable source of information about study participants and the study itself. For instance, carelessness has been linked to participant personality \citep[e.g.,][]{bowling2016,desimone2020} as well as design factors such as study length \citep[e.g.,][]{bowling2021length} and item wording. In particular negatively-worded items are prone to \textit{misresponses} \citep[e.g.,][]{swain2008,weijters2012} triggered by carelessness. Consequently, it is widely recommended to carefully screen survey data for the presence of careless responding \citep[e.g.,][]{desimone2015,huang2015}.

Numerous methods for identifying careless respondents after questionnaire administration (so-called \emph{screeners}) have been proposed, for instance inconsistency indicators such as (resampled) personal reliabilty (\citealp{jackson1976, curran2016}; see also \citealp{goldammer2024}) or psychometric synonyms/antonyms \citep{meade2012}, invariability indicators such as the longstring index \citep{johnson2005}, or multivariate outlier analyses \citep[e.g.,][]{curran2016}. 
We refer to \citet{arthur2021}, \citet{desimone2015}, and \citet{ward2023} for detailed reviews.\footnote{A popular pre-administration method for the detection of careless responding is the inclusion of detection items \citep[e.g., Table~1 in][]{arthur2021}. Such detection items are based on the presumption that an attentive respondent will respond in a specific way, while careless respondent may fail to do so. For instance, it is expected that an attentive respondent would strongly disagree to so-called bogus items such as ``I am paid biweekly by leprechauns'' \citep{meade2012}. Alternative types of detection items are instructed items and self-report items \citep{arthur2021}.} 
More recently, \citet{bloy2025}, \citet{alfons2024}, \citet{ozaki2024}, and \citet{schroeders2022} have studied machine learning techniques for identifying careless respondents.
Existing methods and their implementations in freely available software packages such as \pkg{careless} \citep{careless} make screening for careless respondents a simple task for empirical researchers, thus elevating reliability and validity of research findings and playing an instrumental role in gaining insights into the phenomenon of careless responding. In addition, many popular screeners are intuitive and easy to understand. It should be noted that common indicators typically compute a single score for each participant, with respondents deemed careless and removed from subsequent analysis if their score exceeds a certain cutoff value. 
Although there exist many comparative studies on such screeners \citep[e.g.,][]{curran2016,desimone2018,goldammer2020,huang2012}, their behavior under partially careless responding is unfortunately not well-understood. 
Furthermore, excluding partially careless respondents---particularly in lengthy surveys---may leave a lot of valuable information unused, namely the respective parts of the survey in which they responded attentively.

We therefore aim to address a more nuanced question: Can we identify at which point in the survey participants start responding carelessly, if at all? Indeed, there is evidence that many---though not all---careless respondents tend to begin a survey as attentive and truthful respondents, but may resort to careless responding as the survey progresses, for instance due to fatigue or boredom \citep{bowling2021length, brower2018, galesic2009, gibson2020, ward2017}; cf.~also the literature on back random responding \citep[e.g.,][]{clark2003} or partial random responding \citep[e.g.][]{berry1992, gallen1997partially}. Successful identification of the onset of careless responding can therefore yield valuable, complementary insights to those from existing screeners. For subsequent analysis, the issue of partially careless responding may be transformed into a missing data problem. 
Rather than excluding partially careless respondents altogether, researchers can restrict their primary analyses to the parts of attentive responses, treating the careless responses as missing values. Common guidelines for handling missing data \citep[e.g.,][]{newman2014} can thereby be followed. 
In addition, accurately identifying the onset of careless responding may uncover implications for survey design (e.g., if many participants start responding carelessly at a similar point in the survey, it may be beneficial to rephrase or replace those items, or to place an intervention) and may lead to new theory and a deeper understanding of careless responding behavior (e.g., potentially linking certain personality traits to a tendency for an earlier or later onset of careless responding; cf.~\citealp{gibbert2021}, for theory-building based on outlying data).

For the purpose of identifying at which point in the survey a given participant becomes careless, if at all, we introduce a novel method called \emph{careless onset detection in extensive rating-scale surveys} (CODERS). 
Specifically, CODERS estimates the item at which carelessness onsets in each questionnaire participant, or an absence thereof. An important issue is that careless responding may manifest in multiple distinctive ways \citep[e.g.,][]{curran2016,desimone2015,desimone2018,edwards2019,ward2023}. To capture various types of carelessness, CODERS can combine multiple dimensions of evidence for careless responding to construct a score that, for each item, measures if a given respondent has started responding carelessly by that item. It achieves this by searching for changepoints along a multidimensional series of indices that are potentially indicative of carelessness, such as measures of response inconsistency or invariability.

We argue that the notion of a changepoint is intuitive when studying partial carelessness: if a participant starts responding carelessly, we expect a structural break in their responses. In particular, such a respondent may abandon content-based responding and resort to careless response types from the changepoint onward.
CODERS is highly flexible as it does not assume a statistical model, nor does it require predefining what types of careless response types may exist, and it is primarily intended for lengthy multi-scale surveys. 
Moreover, CODERS builds on nonparametric estimation theory for cutoff values with statistical guarantees regarding the false positive rate (i.e., the probability of attentive respondents being incorrectly flagged as partially careless).
To the best of our knowledge, only one other approach for systematically detecting the onset of careless responding has been proposed to date: The procedure of \citet{yu2019change} is based on item-response-theory (IRT) modeling and is more narroly focused on random responding.

Through extensive simulation experiments, we find that CODERS accurately estimates carelessness onset (or a lack thereof) and achieves excellent sensitivity and specificity in detecting partially careless responding.
Crucially, CODERS should not be viewed as a replacement of existing screeners for careless respondents, but as a complementary tool for a related but more nuanced task, as highlighted by an empirical application.
Re-examining data from a a seminal study on careless responding in Big Five personality measurements \citep{johnson2005}, we demonstrate novel insights that can be obtained by estimating onset items of partially careless responding. 
Reflecting on those insights, we subsequently discuss limitations and potential extensions of CODERS. Our concluding discussion highlights how CODERS complements various existing approaches for the detection of careless respondents.
Finally, we provide the freely available software package 
\pkg{carelessonset} \citep{R:carelessonset}, 
which implements CODERS in the statistical programming environment~\proglang{R} \citep{R} and can be downloaded from \url{https://github.com/mwelz/carelessonset}.


\section*{Introducing CODERS}

The literature generally identifies three distinct ways in which careless responding may manifest: low internal consistency of the given responses (\emph{inconsistency}), low variability of given responses (\emph{invariability}), and impossibly fast responses \citep[e.g.,][]{curran2016,desimone2015,desimone2018,edwards2019,ward2023}.
We expect partially careless responding to result in a change in respondent behavior in at least one of these  dimensions. 
CODERS therefore offers a general framework for searching for a joint changepoint in any combination of measurements of these dimensions of carelessness. While the majority of this paper is focused on inconsistency and invariability, we revisit response times in a later section.
As used here, CODERS thus consists of three main components:
\begin{enumerate}
  \item For each participant, a nonparametric changepoint detection test \citep{shao2010} is applied to a combined series of participant-item level measurements of careless responding.
  \item As a measurement of inconsistency, reconstruction errors (RE) obtained via an auto-associative neural network (autoencoder) \citep{kramer1992} are used. An autoencoder can be viewed as a nonlinear generalization of principal component analysis (PCA), and the (participant-item level) reconstruction errors can be interpreted as the difference between observed responses and what is expected under attentive responding.
  \item As a measurement of invariability, we propose a novel algorithm called LongStringPattern (LSP). LSP can be viewed as a generalization of the famous longstring index of \citet{johnson2005} to the participant-item level, also allowing for recurring patterns involving multiple answer categories.
\end{enumerate}

Regarding the first dimension, inconsistency, the autoencoder learns response patterns that characterize attentive responding \citep[cf.][]{alfons2024}.
We expect that random-like, content-independent responses cannot be learned well and are thus poorly reconstructed by the autoencoder, leading to a changepoint in reconstruction performance. 
Regarding the second dimension, invariability, long sequences of identical responses or fixed response patterns are not expected in well-designed surveys (e.g., using positively and negatively coded items, randomized in some form). 
Hence, we expect a changepoint in response variability once a respondent commences to respond carelessly through straightlining or pattern responding behavior. To obtain a granular estimate of carelessness onset, both reconstruction errors and LSP sequences are measured on the participant-item level so that a changepoint can be detected on the item-level for each participant.

Overall, CODERS attempts to capture evidence for the onset of carelessness by combining multiple indicators that are potentially indicative of such an onset, where the different indicators may capture different manifestations of carelessness. Conversely, if a respondent never becomes careless, we do not expect a changepoint in any dimension. Combining multiple indicators is a generally recommended practice to capture various types of careless responding \citep{goldammer2020, huang2012, meade2012, ward2023}.
We provide a detailed description of our assumptions in Appendix~\ref{app:setup} and additional details on CODERS in Appendix~\ref{app:methodology}.

\subsection*{Quantifying Inconsistency with Autoencoders}
\label{sec:inconsistency}

\citet{ward2023} define internal consistency of responses as patterns that are expected based on theoretical/logical grounds or trends in the data. 
For instance, items that are part of the same construct are expected to correlate strongly in most participants, provided that participants are attentive. 
In contrast, inconsistent careless responding \textit{``generate[s] responses that fail to meet an expected level of consistency''} \citep{ward2023}. 
Respondents may engage in inconsistent careless responding if they attempt to conceal their carelessness, for instance by randomly choosing from all response options \citep{ward2023}. 
For our purposes, we consider the defining characteristic of inconsistent careless responding to be content-independent responses that are approximately randomly chosen from the response options, where the probability of being chosen
may differ between response options, such as preferring extreme response options.

In order to identify inconsistent careless responding, we propose to use the machine learning method of autoencoders  
\citep{kramer1992}, which were originally developed to filter random noise in signal processing applications. An autoencoder is a neural network \citep[e.g.,][]{bishop2006, goodfellow2016} that attempts to reconstruct its input. In other words, the output variables are equal to the input variables. The idea behind reconstructing the input data is to learn the internal structures of the data, thereby separating the signal from random noise that lacks structure. Since we consider the defining characteristic of inconsistent careless responding to be near-randomly chosen responses---which may be viewed as random noise in the context of questionnaire data---we expect autoencoders to perform well in distinguishing between attentive responses and inconsistent careless responses.

To achieve the goal of learning the internal structures of a dataset, autoencoders first express the input data in terms of a low-dimensional representation before reconstructing the input data again from this low-dimensional representation (cf. \citealp{kramer1992}, and Chapter 14.6 in \citealp{goodfellow2016}). The use of such a low-dimensional representation makes autoencoders appealing for questionnaire data, as questionnaires typically contain multiple items to measure each construct of interest, such as attitudes, sentiments, or personality traits. In this context, the autoencoder aims to learn the patterns between the latent constructs in the low-dimensional space. For attentive responses, randomness that can be attributed to taking multiple measurements of the latent constructs is thereby filtered out. Since the amount of randomness from these multiple measurements should be relatively low for a well-designed questionnaire, attentive responses are expected to be reconstructed well from the learned patterns in the low-dimensional space of the latent constructs. Inconsistent careless responses, on the other hand, should contain little to no structure also in this low-dimensional space, and are therefore expected to be reconstructed poorly.

\begin{figure}[!t]
\caption{\textit{Schematic example of a neural network with an autoencoder structure.}}
\centering
\begin{tikzpicture}[shorten >=1pt, scale=0.85, every node/.style={transform shape}]
	\tikzstyle{unit}=[draw, shape=circle, minimum size=1.15cm]
  \tikzstyle{hidden}=[draw, shape=circle, minimum size=1.15cm]
  
  \node[unit, fill=cyan](x0) at (0,3.25){\large $\vect{x}_{1}$};
  \node[unit, fill=cyan](x1) at (0,1.75){\large $\vect{x}_{2}$};
  \node[unit, fill=cyan](x2) at (0,0.25){\large $\vect{x}_{3}$};
  \node[unit, fill=cyan](x3) at (0,-1.25){\large $\vect{x}_{4}$};
  
  \node[hidden, fill=yellow](h10) at (3,4){};
  \node[hidden, fill=yellow](h11) at (3,2.5){};
  \node[hidden, fill=yellow](h12) at (3,1.0){};
  \node[hidden, fill=yellow](h13) at (3,-0.5){};
  \node[hidden, fill=yellow](h14) at (3,-2){};
  
  \node[hidden, fill=orange](h20) at (6,1.75){\large $\vect{\widehat{z}}_{1}$};
  \node[hidden, fill=orange](h21) at (6,0.25){\large $\vect{\widehat{z}}_{2}$};
  
  \node[hidden, fill=yellow](h30) at (9,4){};
  \node[hidden, fill=yellow](h31) at (9,2.5){};
  \node[hidden, fill=yellow](h32) at (9,1.0){};
  \node[hidden, fill=yellow](h33) at (9,-0.5){};
  \node[hidden, fill=yellow](h34) at (9,-2){};
  
  \node[unit, fill=cyan](y0) at (12,3.25){\large $\vect{\widehat{x}}_{1}$};
  \node[unit, fill=cyan](y1) at (12,1.75){\large $\vect{\widehat{x}}_{2}$};
  \node[unit, fill=cyan](y2) at (12,0.25){\large $\vect{\widehat{x}}_{3}$};
  \node[unit, fill=cyan](y3) at (12,-1.25){\large $\vect{\widehat{x}}_{4}$};
  
  \draw[->] (x0) -- (h10);
  \draw[->] (x0) -- (h11);
  \draw[->] (x0) -- (h12);
  \draw[->] (x0) -- (h13);
  \draw[->] (x0) -- (h14);
  
  \draw[->] (x1) -- (h10);
  \draw[->] (x1) -- (h11);
  \draw[->] (x1) -- (h12);
  \draw[->] (x1) -- (h13);
  \draw[->] (x1) -- (h14);
  
  \draw[->] (x2) -- (h10);
  \draw[->] (x2) -- (h11);
  \draw[->] (x2) -- (h12);
  \draw[->] (x2) -- (h13);
  \draw[->] (x2) -- (h14);
  
  \draw[->] (x3) -- (h10);
  \draw[->] (x3) -- (h11);
  \draw[->] (x3) -- (h12);
  \draw[->] (x3) -- (h13);
  \draw[->] (x3) -- (h14);
  
  \draw[->] (h10) -- (h20);
  \draw[->] (h11) -- (h20);
  \draw[->] (h12) -- (h20);
  \draw[->] (h13) -- (h20);
  \draw[->] (h14) -- (h20);
  
  \draw[->] (h10) -- (h21);
  \draw[->] (h11) -- (h21);
  \draw[->] (h12) -- (h21);
  \draw[->] (h13) -- (h21);
  \draw[->] (h14) -- (h21);
  
  \draw[->] (h20) -- (h30);
  \draw[->] (h20) -- (h31);
  \draw[->] (h20) -- (h32);
  \draw[->] (h20) -- (h33);
  \draw[->] (h20) -- (h34);
  
  \draw[->] (h21) -- (h30);
  \draw[->] (h21) -- (h31);
  \draw[->] (h21) -- (h32);
  \draw[->] (h21) -- (h33);
  \draw[->] (h21) -- (h34);
  
  \draw[->] (h30) -- (y0);
  \draw[->] (h31) -- (y0);
  \draw[->] (h32) -- (y0);
  \draw[->] (h33) -- (y0);
  \draw[->] (h34) -- (y0);
  
  \draw[->] (h30) -- (y1);
  \draw[->] (h31) -- (y1);
  \draw[->] (h32) -- (y1);
  \draw[->] (h33) -- (y1);
  \draw[->] (h34) -- (y1);
  
  \draw[->] (h30) -- (y2);
  \draw[->] (h31) -- (y2);
  \draw[->] (h32) -- (y2);
  \draw[->] (h33) -- (y2);
  \draw[->] (h34) -- (y2);
  
  \draw[->] (h30) -- (y3);
  \draw[->] (h31) -- (y3);
  \draw[->] (h32) -- (y3);
  \draw[->] (h33) -- (y3);
  \draw[->] (h34) -- (y3);
  
  \draw (0,4.5) node [black, font=\bfseries]{Input Layer};
  \draw (3,5.25) node [black, font=\bfseries]{Mapping Layer};
  \draw (6,3.5) node [black, font=\bfseries]{Bottleneck Layer};
  \draw (9,5.25) node [black, font=\bfseries]{Demapping Layer};
  \draw (12,4.5) node [black, font=\bfseries]{Output Layer};
\end{tikzpicture}
\caption*{\textit{Note.} 
The network consists of five fully connected layers and is symmetric around the central layer.
Each dimension is represented by one node, and nodes within the same layer are aligned vertically. 
In each node, a so-called activation function is applied to a linear combination of the nodes from the previous layer.
The first layer (input layer) represents the input variables (here $\vect{x}_{1}, \dots, \vect{x}_{4}$).
The second layer (mapping layer) prepares the data to be compressed in the third and central layer (bottleneck layer), which holds low-dimensional representations of the input data (here $\vect{\widehat{z}}_{1}$ and $\vect{\widehat{z}}_{2}$). 
The fourth layer (demapping layer) prepares the low-dimensional representations for reconstructing the input data in the final layer (output layer), yielding reconstructions in the original dimension (here $\vect{\widehat{x}}_{1}, \dots, \vect{\widehat{x}}_{4}$).
}
\label{fig:autoencoder-scheme}
\end{figure}

A schematic example of the network architecture of an autoencoder is given in Figure~\ref{fig:autoencoder-scheme}. Concretely, an autoencoder is a fully connected network whose nodes are organized in multiple layers. Each node takes a linear combination of the nodes from the previous layer and subsequently applies a so-called activation function in order to allow for nonlinear relationships. A crucial part of an autoencoder is the central layer, known as the bottleneck layer, which stores the low-dimensional representations of the input data. The network architecture of an autoencoder is symmetric around this bottleneck layer such that it can be viewed as having an encoder-decoder architecture: the encoder compresses the data into the low-dimensional representations, while the decoder reconstructs the input data again. It follows that an autoencoder can be seen as a dimension reduction technique, specifically as a nonlinear generalization of principal component analysis \citep[PCA; see][]{kramer1991}.

Hence, the number of nodes in the bottleneck layer is an important design choice in the architecture of an autoencoder. In the context of questionnaire data, we recommend setting this number equal to the number of constructs measured in the questionnaire. For example, consider data from the Revised NEO Personality Inventory \citep[NEO-PI-R;][]{costa1992revised}, which contains 240 items and measures six so-called facets of each of the Big Five personality traits. Thus, this instrument measures a total of 30 facets, such as anxiety and modesty, so that we would set the number of nodes in the bottleneck layer to~30.

Further intuition and a thorough mathematical definition of autoencoders can be found in Appendix~\ref{app:methodology}.
Although there are other design choices besides the number of nodes in the bottleneck layer, our software implementation in package 
\pkg{carelessonset} \citep{R:carelessonset} 
provides default values that we recommend researchers to use. 
The only input required by the user is the number of constructs measured in the survey (for the number of nodes in the bottleneck layer).

\begin{figure}[!t]
\caption{\textit{Simulated illustrative example of autoencoder reconstruction errors of an inconsistent partially careless respondent to a 300-item questionnaire comprising five-point Likert scales.}}
\centering
\includegraphics[width = 0.8\textwidth]{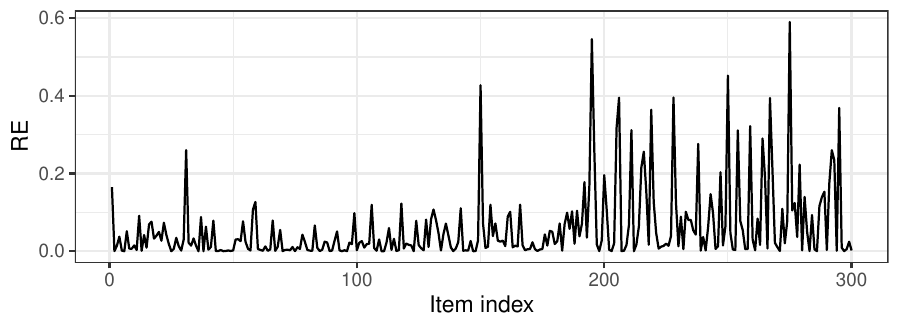}
\caption*{\textit{Note.} The simulated participant started choosing answer categories completely at random from the 183rd item onward.}
\label{fig:exampleRE}
\end{figure}

Suppose now that we have obtained autoencoder reconstructions of the responses of each survey participant. 
Let there be~$n$ participants and~$p$ items, and denote the observed response of the~$i$-th participant to the~$j$-th item by~$x_{ij}$ and the corresponding reconstruction by~$\widehat{x}_{ij}$, for $i=1,\dots,n$ and $j=1,\dots, p$.  
We define the reconstruction error~$\textsf{RE}_{ij}$ associated with the response of participant~$i$ to item~$j$ to be the squared difference between the observed response and its reconstruction, scaled by the range of the answer categories. 
Formally, for participants $i=1,\dots,n$ and items $j=1,\dots,p$,
\begin{equation} \label{eq:RE}
\textsf{RE}_{ij} = \left(\frac{x_{ij} - \widehat{x}_{ij}}{L_j - 1} \right)^2,
\end{equation}
where~$L_j$ denotes the number of answer categories of item~$j$. Recall that we expect the reconstruction errors~$\textsf{RE}_{ij}$ to be low in case of attentive responding but high in the presence of inconsistent careless responding. 
Hence, if participant~$i$ starts providing inconsistent careless responses at item~$k\in\{1,\dots,p\}$, we expect a changepoint  at position~$k$ in the participant's series of~$p$ reconstruction errors, as reconstruction errors are expected to be higher from item~$k$ onward. An example can be found in Figure~\ref{fig:exampleRE}, where we show simulated responses of a hypothetical participant in a 300-item questionnaire (5 answer categories per item) who starts choosing answer categories completely at random from the 183rd item, resulting in systemically larger autoencoder reconstruction errors from that item onward.

\subsection*{Quantifying Invariability}

We propose to measure response invariability through a novel algorithm inspired by the longstring index of \citet{johnson2005} that we call LongStringPattern (LSP). This algorithm exploits that invariable careless responses are characterized by content-independent response patterns. A pattern sequence has the defining property that it consists of repeated occurrences of the same response pattern. 
Consider the sequence of responses $1-2-1-2-\ldots-1-2$, which consists of repeated occurrences of the pattern $1-2$. 
Denoting the length of a pattern by~$l$, the above example exhibits a pattern of length~$l=2$.
Analogously, the sequence $1-2-3-1-2-3-\ldots-1-2-3$ consists of repeated occurrences of the pattern $1-2-3$ of length~$l=3$, whereas the straightlining sequence $1-1-\ldots-1$ consists of repeated occurrences of~1, which is a pattern of length~$l=1$.

Define $l\textsf{-pattern}_{ij}$ as the number of consecutive items in the longest sequence of a recurring pattern of length~$l$ in which the response of participant~$i$ to item~$j$ takes part, or assign the value~1 if the response is not part of a recurring pattern of length~$l$.
Consider the following two illustrative examples. First, for~$l=2$, the response sequence $1-2-1-2-1-4-3-5-4-5-4$ will be assigned the 2-pattern sequence $4-4-4-4-1-1-1-4-4-4-4$ because the first four responses and last four responses comprise two occurrences of the distinct patterns $1-2$ and $5-4$, respectively, which are both of length~$l=2$. 
The central three responses ($1-4-3$) are not part of any recurring pattern of length~$l=2$, hence they are each assigned the value~1. 
Second, for a recurring pattern of length $l=1$ (i.e., consecutive identical responses), the response sequence $3-2-3-3-1-4-1-1-1$ is assigned the 1-pattern sequence $1-1-2-2-1-1-3-3-3$, since the subsequences of consecutive identical responses $3-3$ and $1-1-1$ comprise two and three responses, respectively.\footnote{The 1-pattern sequence can be seen as a generalization of the the longstring index of \citet{johnson2005}, which we recover by calculating $\max_{1 \leq j \leq p} 1\textsf{-pattern}_{ij}$ for each respondent $i = 1, \dots, n$.} 
In short, higher values in the~$l$-pattern sequence capture a tendency to respond in fixed patterns of length~$l$. If a participant starts engaging in invariable careless responding, we expect a changepoint in the~$l$-pattern sequence (for some~$l$) from relatively low to relatively high values, assuming that the survey design precludes such patterns in attentive responses.

However, an $l$-pattern sequence crucially depends on the choice of pattern length~$l$, which is restrictive since invariable careless respondents may choose widely different patterns. 
Consequently, a single choice of pattern length~$l$ is unlikely to capture all careless response patterns that may emerge. 
To tackle this issue, we propose to calculate the~$l$-pattern sequence for multiple choices of~$l$, and we define
\begin{equation} \label{eq:LSP}
\textsf{LSP}_{ij} = \max_{l} \ l\textsf{-pattern}_{ij}
\end{equation}
for participant $i = 1, \dots, n$ and item $j = 1, \dots, p$. We call the series of within-participant values $\textsf{LSP}_{ij}$ across all items a LongStringPattern (LSP) sequence. In this paper, we vary~$l$ from 1 to the number of answer categories in the survey.
The rationale behind this choice is that a careless respondent may repeat, e.g., the pattern $1-2-3-4-5$ in case of five answer categories. 
At the same time, we consider it unlikely that careless respondents choose complicated individual patterns whose length exceeds the number of answer categories. 

\begin{figure}[!t]
\caption{\textit{Simulated illustrative example of a LongStringPattern sequence of an invariable partially careless respondent to a 300-item questionnaire comprising five-point Likert scales.}}
\centering
\includegraphics[width = 0.8\textwidth]{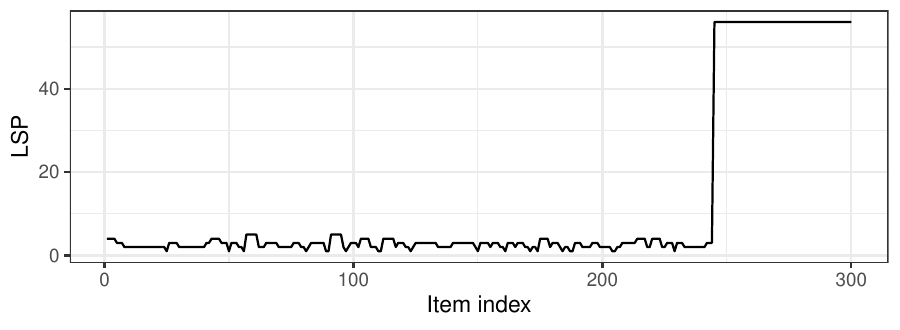}
\caption*{\textit{Note.} The simulated participant started responding carelessly with the recurring pattern $1-3$ from the 245th item onward.}
\label{fig:exampleLSP}
\end{figure}

We provide a detailed description of the LSP algorithm in Appendix~\ref{app:methodology}.
An example of an LSP sequence is shown in Figure~\ref{fig:exampleLSP}, which shows simulated responses of a hypothetical participant in a 300-item questionnaire (5 answer categories per item) who starts responding carelessly using the recurring pattern $1-3$ from the 245th item, resulting in systematically larger LSP values from that item onward.

\subsection*{Identifying Carelessness Onset via Changepoint Detection}

For each participant, we can obtain multiple individual series of length equal to the total number of items in the survey. Each individual series should thereby measure one of the dimensions indicative of carelessness. Although we focus on using reconstruction errors from~\eqref{eq:RE} to measure inconsistency as well as the LSP sequence from~\eqref{eq:LSP} to measure invariability, 
in principle CODERS can also be used with any single measurement or any other combination of different measurements of carelessness on the participant-item level.

As motivated in the previous two sections, we expect a changepoint in any of the measurements of carelessness from the onset of careless responding (cf.~Figures~\ref{fig:exampleRE} and~\ref{fig:exampleLSP}), and no changepoint in the absence of carelessness. 
We therefore propose to apply the nonparametric cumulative sum self-normalization test of \citet{shao2010}, which was developed in the context of time series analysis. It searches for the location of a possible changepoint in a multivariate series and has attractive theoretical guarantees (see \citealp{shao2010, zhao2022}; and our detailed description in Appendix~\ref{app:methodology}).

For each point in a multivariate series (we use measurements of carelessness associated with each questionnaire item), this changepoint detection test calculates a statistic that compares a quantity of interest (we take the mean) in the periods before and after the current point (item). 
If the maximum of this statistic exceeds a critical value, the point (item) where this maximum occurs is flagged as a changepoint. Otherwise no changepoint is flagged. 
The critical value thereby depends only on a pre-specified significance level~$\alpha$ and the dimension of the multivariate series. 
Importantly, it controls the size of the test regardless of the distribution of the individual series. 
This gives CODERS great flexibility, as it allows us to use any combination of measurements of carelessness without making possibly unrealistic assumptions on their distributions or obtaining cutoff values in an ad-hoc manner. 
In line with literature that suggests that there should be overwhelming evidence when labeling respondents as careless \citep[e.g.,][]{huang2012}, we recommend using a conservative significance level no greater than~1\% for identifying changepoints.

\begin{figure}[!t]
\caption{\textit{Illustrative example output from CODERS using responses of two simulated partially careless respondents to a 300-item questionnaire.}}
\centering
\begin{subfigure}{0.49\textwidth}
\centering
\includegraphics[width = \textwidth]{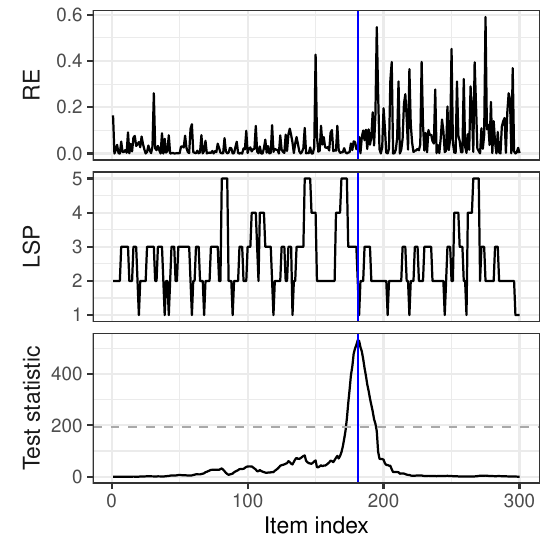}
\caption{Inconsistent partial carelessness.}
\label{fig:examples-inconsistent}
\end{subfigure}
\begin{subfigure}{0.49\textwidth}
\centering
\includegraphics[width = \textwidth]{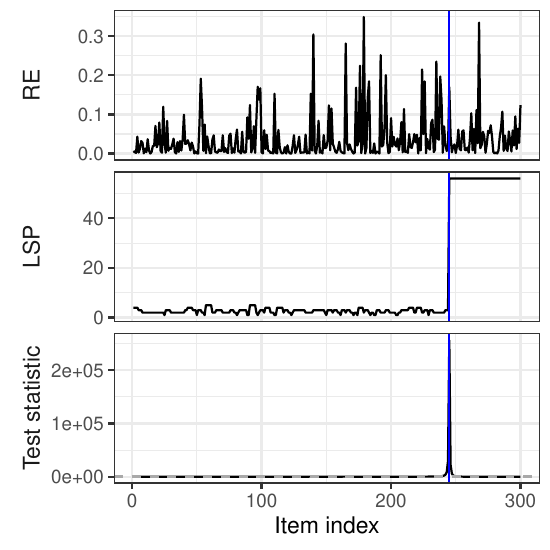}
\caption{Invariable partial carelessness.}
\label{fig:examples-invariable}
\end{subfigure}
\caption*{\textit{Note.} The left subfigure shows the output from CODERS for a simulated partially inconsistent respondent who starts responding randomly from the 183rd item onward, while the right subfigure does so for a simulated partially invariable respondent who starts repeating the pattern $1-3$ from the 245th item onward. Each subfigure visualizes the autoencoder reconstruction errors (RE), LongStringPattern (LSP) sequence, and test statistics associated with the corresponding participant's responses. The blue solid vertical lines indicate the carelessness onset items identified by CODERS at significance level~0.1\%, which correspond to item~181 (two items before true onset) for the partially inconsistent respondent, and to the true onset item~245 for the partially invariable respondent. The dashed horizontal line is the critical value of the test statistic at the 0.1\% significance level (192.5; see Table~\ref{tab:critvals} in Appendix~\ref{app:methodology}). The reconstruction errors in the left subfigure are those shown in Figure~\ref{fig:exampleRE}, and the LSP sequence in the right subfigure is that shown in Figure~\ref{fig:exampleLSP}.}
\label{fig:examples}
\end{figure}

As an illustrative example, consider Figure~\ref{fig:examples-inconsistent}. It illustrates two dimensions computed from a given participant's responses, namely autoencoder reconstruction errors and the LSP sequence. 
The former series has been shown before in Figure~\ref{fig:exampleRE}.
The bottom panel of Figure~\ref{fig:examples-inconsistent} shows the test statistics computed from this two-dimensional series, where the maximum value at the 181st item clearly exceeds the critical value. 
That is, CODERS identifies inconsistent carelessness from the 181st item onward, as the reconstruction errors are in general much larger than before. 
Indeed, this simulated participant started choosing responses at random just two items after the estimated onset.  
Figure~\ref{fig:examples-invariable} is similar, but computed from a different respondent, whose LSP sequence has been shown before in Figure~\ref{fig:exampleLSP}. CODERS flags a changepoint at the 245th item. From this item onward, the LSP values are clearly elevated, indicating invariable carelessness. In fact, this simulated participant started repeating the pattern $1-3$ at the identified item.


\section*{Simulation Experiments}\label{sec:simulation}

After introducing CODERS, we now investigate its behavior via simulation experiments modeled after data collected by \citet{johnson2005}. 
We use the data from this seminal study on careless responding in our empirical application in a later section, where the data are described in more detail.
Unfortunately, it is not possible to compare CODERS with the approach of \citet{yu2019change}, which is the only other method for detecting the onset of careless responding that we are aware of, as no \proglang{R} code seems to be available.

\subsection*{Data Generation}

\begin{figure}[!t]
\caption{\textit{Correlation matrix and marginal response distributions used in simulations.}}
\centering
\begin{subfigure}[t]{\textwidth}
\centering
\includegraphics[width = 0.75\textwidth]{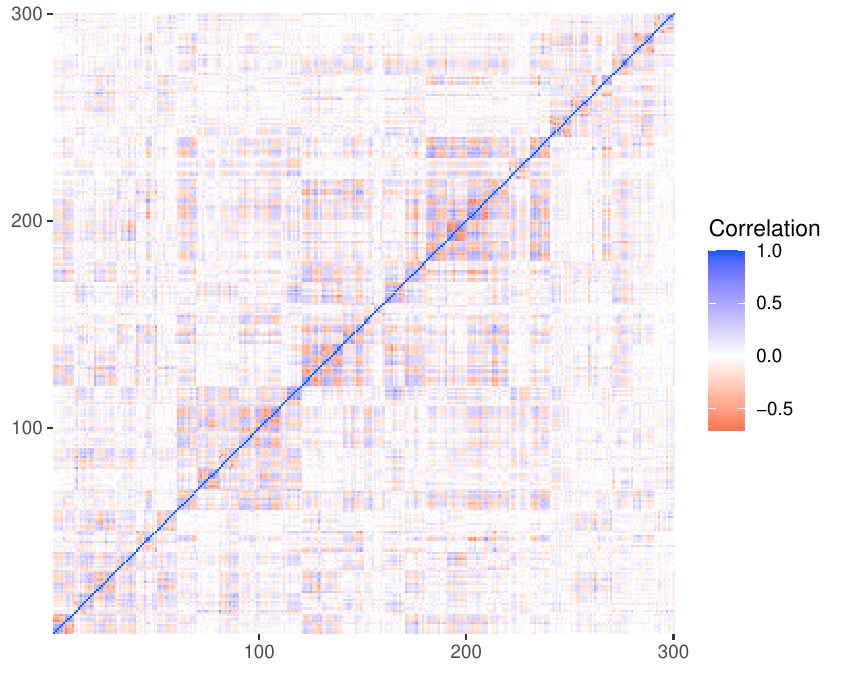}
\caption{Correlation matrix of the $p=300$ items. A more saturated blue (red) in a cell corresponds to a stronger positive (negative) correlation between the corresponding items.}
\end{subfigure}
\begin{subfigure}[t]{\textwidth}
\centering
\includegraphics[width = \textwidth]{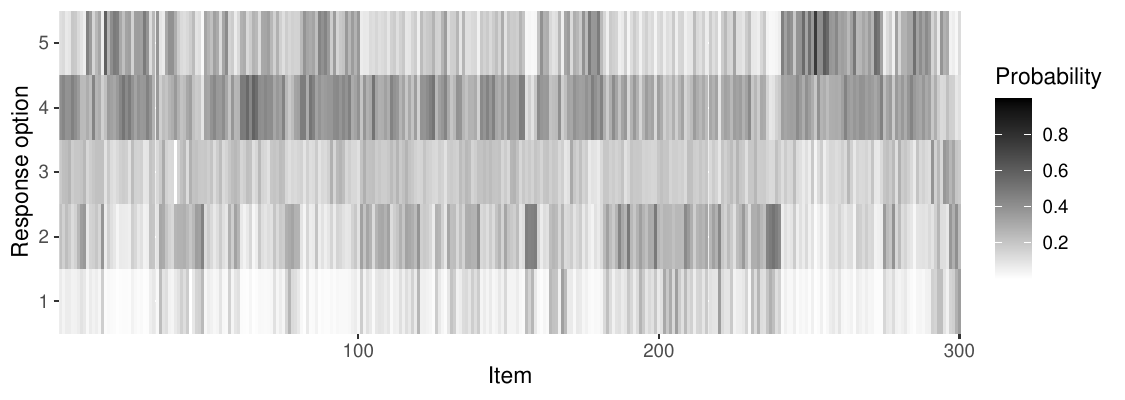}
\caption{Marginal response distributions of the $p=300$ items ($x$-axis), each of which has five response options ($y$-axis). A darker shade of gray indicates a higher probability of the corresponding response option.}
\label{fig:marginals}
\end{subfigure}
\caption*{\textit{Note.} 
The correlation matrix and the marginal response distributions are calculated from the data of \citet{johnson2005} after removing presumably careless respondents. Of the $p=300$ items, 152 are positively scored and 148 are negatively scored. The items in both plots are presented in an order such that items measuring the same facet are adjacent. 
}
\label{fig:cormat-marginals}
\end{figure}

The survey instrument used by \citet{johnson2005} is a 300-item representation of the NEO-PI-R \citep{goldberg1999} measurement of the Big Five personality traits \citep[openness, conscientiousness, extraversion, agreeableness, and neuroticism;][]{goldberg1992} by means of five-point Likert scales.
Each of the five traits comprises six subdomains (facets) so that a total of $5\times 6 = 30$ facets are measured. 
We base our simulation design on the between-item correlation structure and per-item response distributions in the data collected by \citet{johnson2005}, calculated after removing any presumably careless respondents detected in our empirical application.  
Figure~\ref{fig:cormat-marginals} visualizes the resulting correlation matrix and marginal response distributions. Of the 300 items, 152 are positively scored while 148 are negatively scored.
In the correlation matrix, one can identify the five traits and, within each block of within-trait correlations, also the six corresponding facets. 
Using these correlations and marginal response distributions, we generate rating-scale responses of $n=500$ participants to the $p=300$ items,\footnote{We use package \pkg{simstudy} \citep{goldfeld2020} to generate the data, for which the target values of correlations only hold approximately on the population level.} with the items in the same randomized order as in \citet{johnson2005}.

Subsequently, a fixed percentage  $\gamma \in \{0, 0.02, 0.04, 0.1, 0.2, 0.3, 0.4, 0.5, 0.6, 0.8, 1\}$ of \mbox{respondents} is selected to be partially careless. That is,~$\gamma$ denotes the prevalence of careless responding. For each of the selected partially careless respondents, we sample a carelessness onset item from which onward the remaining responses are careless. 

We consider three regimes for the carelessness onset: (i) the \emph{baseline} setting where the onset item is randomly sampled from  $\{30,31,\dots,269,270\}$ so that carelessness starts between 10\% and 90\% of all items, (ii) \emph{early onset} randomly sampled from items $\{30,31,\dots,149,150\}$ (between 10\% and 50\% of all items), and (iii) \emph{late onset} randomly sampled from items $\{150,151,\dots,269,270\}$ (between 50\% and 90\% of all items).

From the carelessness onset item onward, we generate the responses from one of the following four careless response types: (i) \emph{random responding} (choosing answer categories completely at random), (ii) \emph{extreme responding} (randomly choosing between the two most extreme answer categories), (iii) \emph{straightlining} (repeating the same, randomly determined answer category), or (iv) \emph{pattern responding} (repeating a fixed pattern such as $1-2-3$ or $5-4$, which is randomly determined from all possible patterns of length two to five). 
Specifically, we assign these careless response types to the partially careless respondents so that there are \mbox{$\lfloor \gamma n / 4\rfloor$} respondents for each response type, where $\lfloor \ . \ \rfloor$ denotes the operator for rounding down to the nearest integer.

We repeat the above described data generating process~100 times. 
We compare three variations of CODERS based on (i)~both reconstruction errors (RE) and the LSP sequence, (ii)~RE only, and (iii)~the LSP sequence only. For the autoencoder, we set the number of nodes in the bottleneck layer equal to the number of measured constructs (i.e.,~30 facets). 
In all three variations of CODERS, we look for changepoints with three different significance levels~$\alpha$, namely 0.1\%, 0.5\%, and 1\% (from most to least conservative).

\subsection*{Performance Evaluation}

To quantify how accurately partially careless respondents are detected, we compute the false positive rate (FPR) (proportion of true attentive respondents incorrectly flagged as careless) and the false negative rate (FNR) (proportion of true careless respondents incorrectly \emph{not} flagged as careless). 
For CODERS, we consider a respondent to be flagged as careless if a changepoint is detected. 
In addition, to quantify how accurately the onset of careless responding is detected, we compute the mean absolute error (MAE) (difference between the true onset item and the detected onset item in absolute value, averaged over the set of true careless respondents for whom a changepoint is detected). 
We report averages of these performance measures across the~100 simulation runs. 
For all three measures, a lower value (close to 0) indicates better performance.

\subsection*{Results}

\begin{figure}[!t]
 \caption{\textit{False positive rate of CODERS for attentive respondents.}}
 \centering
  \includegraphics[width = \textwidth]{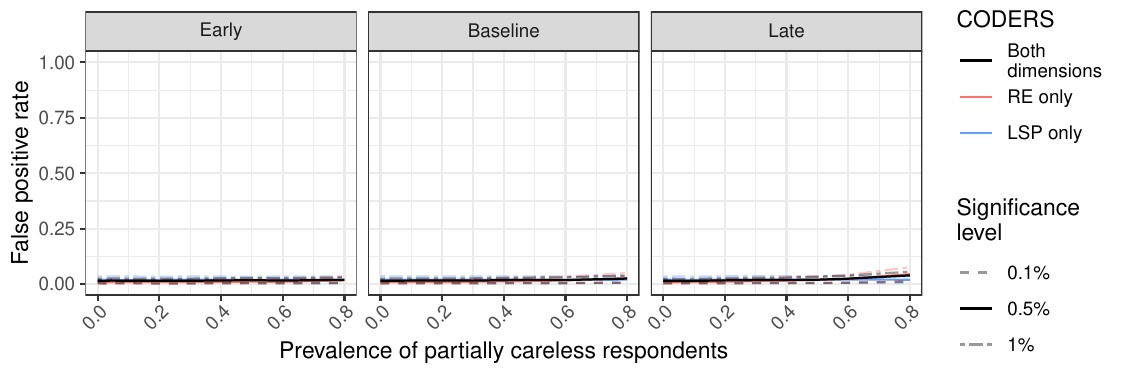}
  \caption*{\textit{Note.} The $x$-axis displays various prevalence levels of partially careless respondents. Results for different onset regimes are shown in separate columns, averaged across 100 repetitions.}
  \label{fig:CODERS-main-fpr}
\end{figure}

We focus on discussing the results for CODERS using both measurements of carelessness (RE and LSP), except when we explicitly state otherwise.
Varying the prevalence of careless responding, Figure~\ref{fig:CODERS-main-fpr} visualizes the FPR of CODERS in fully attentive respondents for the different carelessness onset regimes (in separate columns).
Overall, CODERS rarely flags changepoints in fully attentive respondents. Even with our most liberal choice of significance level~$\alpha$ (1\%), overrejection increases only slowly to about 3\% for 60\% prevalence, before it reaches between 3.0\% and 5.6\% at 80\% prevalence (depending on the onset regime).
Other than for extremely high prevalence levels, the FPR stays reasonably close to the respective nominal level, indicating that CODERS controls Type~I errors fairly well.

\begin{figure}[!t]
\caption{\textit{Performance measures of CODERS for partially careless respondents.}}
\centering
\includegraphics[width = \textwidth]{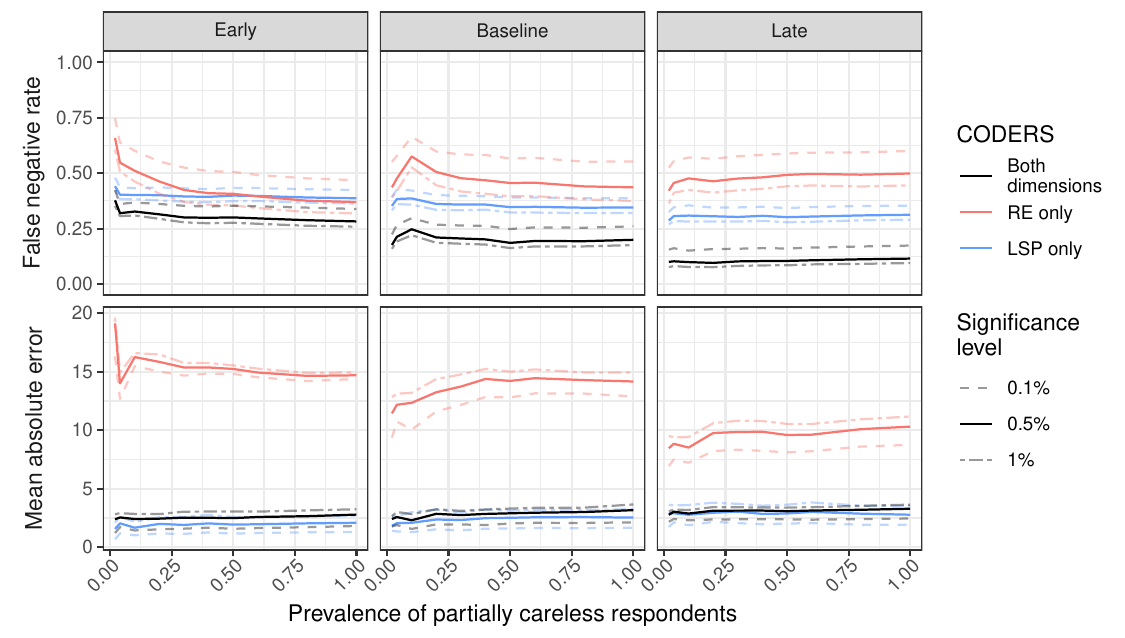}
\caption*{\textit{Note.} The $x$-axis displays various prevalence levels. The top row shows results for the false negative rate, while the bottom row visualizes the accuracy of onset estimation via the mean absolute error. Results for different onset regimes are shown in separate columns, averaged across 100 repetitions.}
\label{fig:CODERS-main-careless}
\end{figure}

For partially careless respondents, Figure~\ref{fig:CODERS-main-careless} (top row) contains a similar plot of the FNR. 
Most importantly, the FNR of CODERS with both dimensions remains remarkably constant throughout prevalence levels, at about 18--26\% in the baseline onset regime depending on the significance level. 
While the FNR increases to about 36\% with the most conservative significance level (0.1\%) in the early onset regime, it decreases to about 8\% with the most liberal significance level (1\%) in the late onset regime. 
Overall, the relatively low FNR implies that CODERS successfully identifies a large majority of the true careless respondents.
Combined with its low FPR, we find that CODERS discriminates well between attentive and partially careless respondents.

In addition, Figure~\ref{fig:CODERS-main-careless} (bottom row) displays the MAE of the estimated onset in partially careless respondents. 
The prevalence level and onset regime have little effect on the accuracy of the estimates, with an MAE hovering around 2 to 3, depending on the significance level. 
That is, on average the estimated onset is only two to three items away from the true onset, indicating that CODERS achieves excellent accuracy in estimating carelessness onset.

\begin{figure}[!t]
\caption{\textit{False negative rate of CODERS for different types of partially careless respondents.}}
\centering
\includegraphics[width = \textwidth]{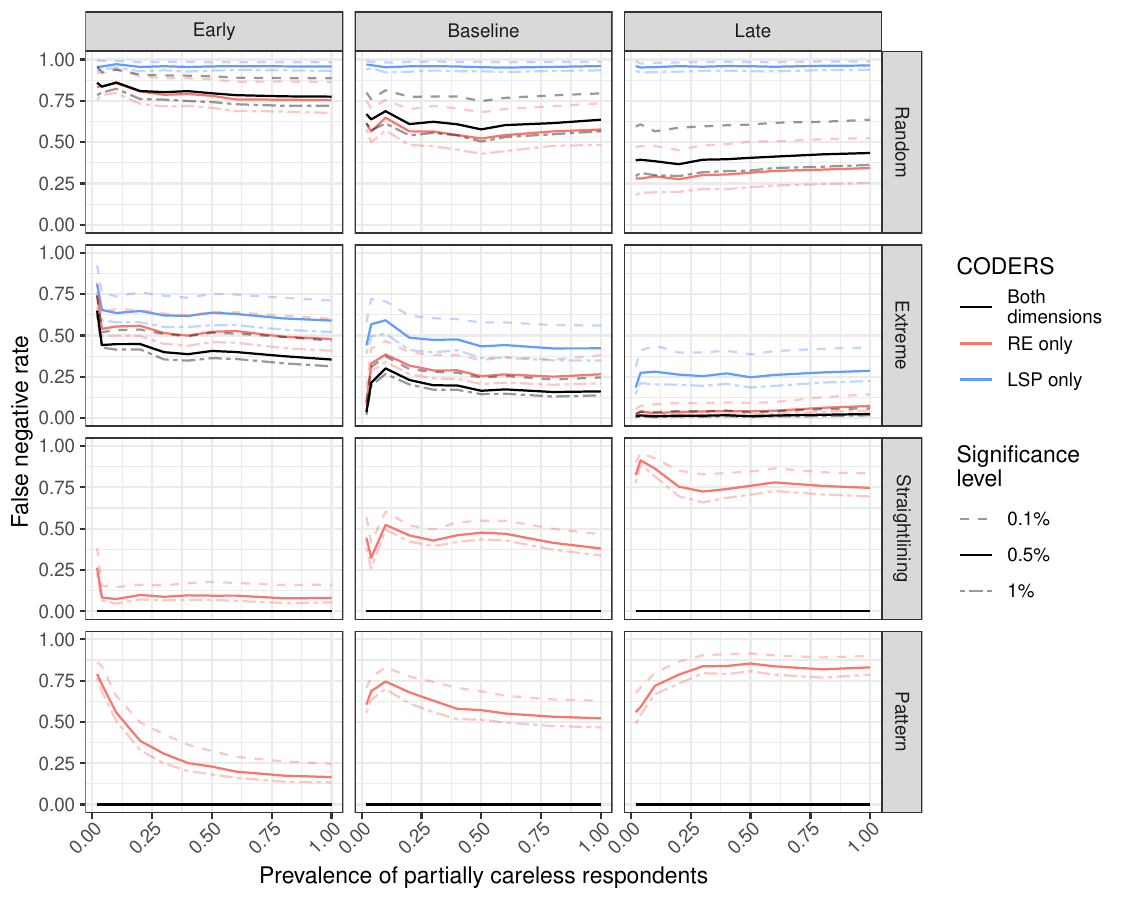}
\caption*{\textit{Note.} The $x$-axis displays various prevalence levels. Results for different types of careless responding (rows) in different onset regimes (columns) are averaged across 100 repetitions.}
\label{fig:CODERS-main-fnr}
\end{figure}

\begin{figure}[!t]
\caption{\textit{Mean absolute error of carelessness onset estimation with CODERS for different types of partially careless respondents.}}
\centering
\includegraphics[width = \textwidth]{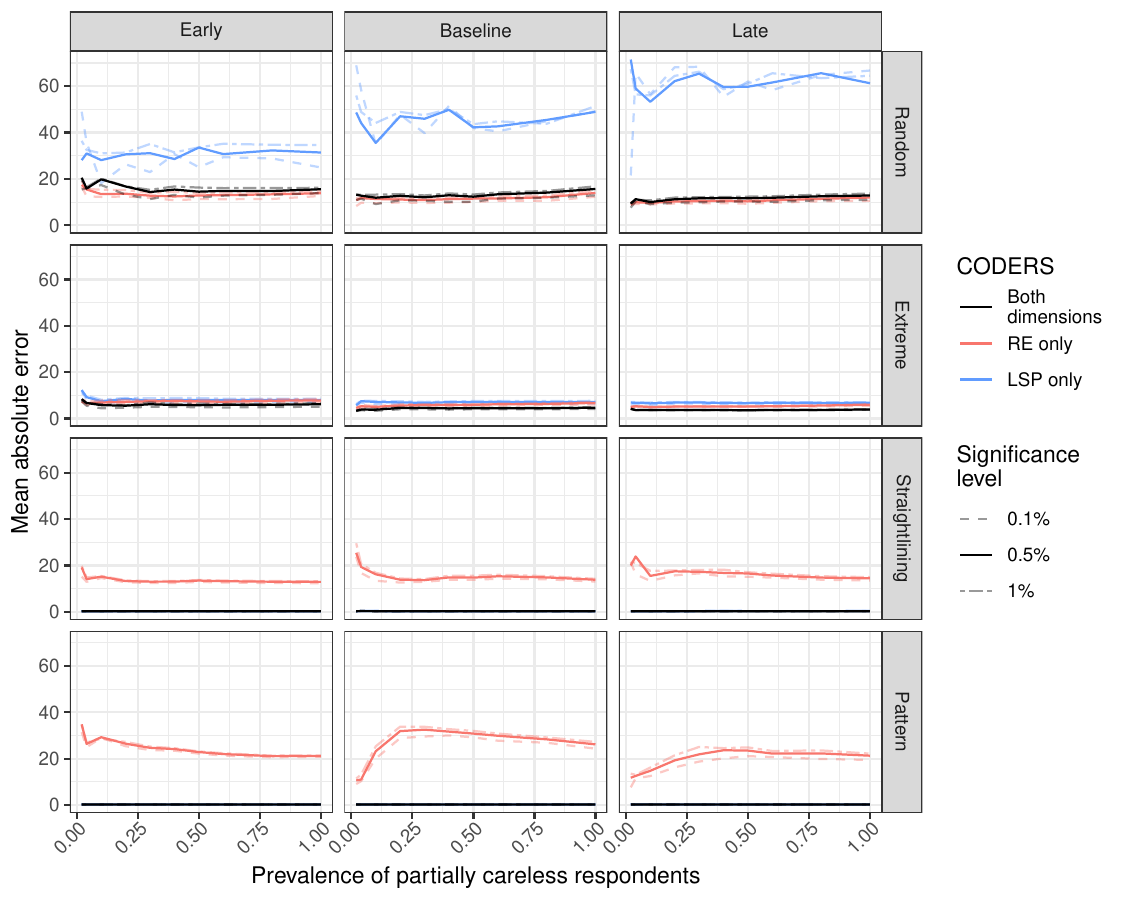}
\caption*{\textit{Note.} The $x$-axis displays various prevalence levels. Results for different types of careless responding (rows) in different onset regimes (columns) are averaged across 100 repetitions.}
\label{fig:CODERS-main-mae}
\end{figure}

On the other hand, Figure~\ref{fig:CODERS-main-careless} further reveals that the FNR of CODERS using only RE or LSP is considerably higher in all settings.
Furthermore, the MAE of the estimated onset is much higher when using only RE.
It is important to note that such a drop in performance is expected when using only one measurement of carelessness due to the various types of careless responding being simulated.
RE is designed to capture inconsistent carelessness and may therefore miss invariable carelessness, whereas it is the other way around for LSP. 
To investigate this further, Figures~\ref{fig:CODERS-main-fnr} and~\ref{fig:CODERS-main-mae} show the FNR and the MAE of onset estimation averaged over subgroups of participants corresponding to each type of careless responding (in separate rows). For the two inconsistent types of carelessness (top two rows), using both dimensions yields overall good results similar to to those using only RE, whereas performance is clearly worse when using only LSP. For the two invariable types of carelessness (bottom two rows), the results when using both dimensions are near-perfect (with FNR and MAE close to 0) and indistinguishable from using only LSP, while performance deteriorates when using only RE. Hence, these findings highlight the benefit of combining measurements of different dimensions of carelessness in CODERS.

We now zoom in on the results of CODERS for the two types of inconsistent carelessness in Figures~\ref{fig:CODERS-main-fnr} and~\ref{fig:CODERS-main-mae}, as there is notable heterogeneity. Specifically, CODERS performs clearly better for extreme responding than for random responding, with the FNR and the MAE of onset estimation in general being lower. Considering that the items exhibit a variety of different probability distributions for the attentive responses, with a low probability mass in at least one of the extreme answer categories (see Figure~\ref{fig:cormat-marginals}), distinguishing random responding is simply a much more challenging problem than distinguishing extreme responding.

For random responding (see the first row in Figure~\ref{fig:CODERS-main-fnr}), CODERS only achieves a decent FNR in the late onset regime with significance level 1\% or 0.5\% (about 32\% and 40\%, respectively).
In the baseline onset regime, its FNR is higher than 50\%. In the early onset regime, even our most liberal significance level of 1\% only yields an FNR of about 75\%. Nevertheless, this finding should not be mistaken as an intrinsic inability of CODERS to detect random responding. 
As the results greatly improve the later the onset of carelessness, it seems that the autoencoder needs enough attentive responses to learn the challenging task of distinguishing random responses. 
In other words, this is an issue of statistical power, which explains why it can be mitigated to some extent by choosing a less conservative significance level.
For additional investigation, we apply traditional methods for identifying careless respondents also used by \citet{johnson2005}. They do not improve upon CODERS (see Appendix~\ref{app:simulations}), therefore further highlighting the difficulty of detecting random respondents in this simulation design.
Coming back to CODERS, it shows similar but far less pronounced behavior for extreme responding (see the second row in Figure~\ref{fig:CODERS-main-fnr}). The FNR remains decent in the early onset regime, while it is good in the baseline onset regime and excellent (close to 0) in the late onset regime.

To summarize our findings, CODERS performs well for extreme responding and near-perfectly for the two types of invariable responding. 
On the other hand, it struggles with random responding in this simulation design, where decent power is only achieved in case of late carelessness onset and with a not-too-conservative significance level. 
However, results for traditional methods (see Appendix~\ref{app:simulations}) suggest that distinguishing random responding from attentive responding may be an intrinsically difficult problem for the chosen correlation structure and marginal distributions. 
Overall, we find that (i) CODERS reliably discriminates between attentive and partially careless respondents, as indicated by near-zero false positive rates and low overall false negative rates, (ii) CODERS accurately estimates the location of carelessness onset, and (iii) the performance of CODERS is stable across a wide range of prevalence levels, demonstrating robustness.

\subsection*{Additional Simulations}

We conducted additional simulation experiments to further investigate the robustness of our findings and to obtain a better understanding of possible limitations of CODERS.
While details and results can be found in Appendix~\ref{app:simulations}, we provide a brief outline of each experiment and the most relevant findings below.

First, we vary the number of items to investigate the performance of CODERS in shorter surveys. 
Recall that our simulation design is based on measurements of the Big Five personality traits, with each trait being measured by 60 items (for a total of 300 items). 
We randomly select traits in a stepwise manner and drop all items of the selected traits, resulting in surveys of length $p \in \{240, 180, 120, 60 \}$. 
The overall FNR in partially careless respondents increases only slowly as survey length decreases. 
For $p = 60$, only very few random respondents are detected, but the FNR remains decent for extreme responding and near-perfect for the invariable types of carelessness. 
Furthermore, the accuracy of the estimated onset of carelessness stays excellent throughout the investigated survey lengths. 
In summary, other than for random respondents, overall power remains adequate when using a not-too-conservative significance level (e.g.,~0.5\% or~1\%).

Second, we vary the number of participants by setting $n = 248$ and $n = \text{1,000}$.\footnote{Since we generate four types of partially careless respondents in the simulations, the sample size should be divisible by 4, explaining the unusual choice of $n = 248$ (rather than $n = 250$).} 
The overall FNR in partially careless respondents slightly increases for the smaller sample size and slightly decreases for the larger sample size, with bigger changes being observed for random respondents. 
That is, the effect of the sample size on power is relatively small except for random responding, where further improvements in performance may be possible for even larger sample sizes.

Third, we consider an additional type of careless responding to investigate the performance of CODERS when the lines between attentive and careless responding are blurry. 
This type of carelessness, to which we refer as \emph{middling}, is characterized by randomly choosing from the three middle answer categories (out of five).
Given the correlation structure and marginal distributions of attentive responding (see Figure~\ref{fig:cormat-marginals}), there is plenty of natural variation among the central response categories, thus making middling extremely hard to distinguish. 
Indeed, although CODERS maintains a low FPR in fully attentive respondents, it detects only very few middling respondents (as do the traditional methods). 
This further highlights that CODERS requires (inconsistent) partially careless responding to be sufficiently distinct from attentive responding to successfully distinguish between them.

Fourth, we emulate a situation of temporary carelessness with two changepoints: the first constitutes a change from attentive to careless responding, and the second embodies a change back to attentive responding.
We find that CODERS fails to  detect such temporarily careless respondents, but the same applies to the traditional methods (except for the longstring index, which successfully captures temporary straightliners). 
This highlights that while CODERS is effective in detecting a single changepoint, it is simply not designed for detecting multiple changepoints.


\section*{Empirical Application}\label{sec:application}

After studying the performance of CODERS in simulations inspired by the data from \citet{johnson2005}, we now re-examine the responses from this seminal study for partial carelessness.

\subsection*{Data}

The data collected by \citet{johnson2005} concern the five factor model of personality \citep{goldberg1992}, which assumes that variation in a measurement of personality can be explained by the so-called \emph{Big Five} personality traits (factors): openness, conscientiousness, extraversion, agreeableness, and neuroticism. Each of the Big Five factors can be further decomposed into six subdomains (facets): for instance, depression and anxiety are facets of the neuroticism trait. Consequently, common Big Five instruments, such as the 240-item NEO-PI-R \citep{costa1992revised}, typically measure a total of~$5\times 6 = 30$ facets. The data set of \citet{johnson2005} consists of internet-collected responses to a 300-item representation of the NEO-PI-R inventory \citep{goldberg1999}, in which each of the~30 facets is measured by~10 items on a five-point Likert scale.

\citet{johnson2005} identifies straightliners via the \emph{longstring index} (which the author proposed in that paper) and inconsistent careless respondents via \emph{personal reliability} \citep{jackson1976} and \emph{psychometric antonym} \citep[e.g.,][]{meade2012} scores. The data from this study are made publicly available by the author at \url{https://osf.io/sxeq5/}. Importantly, from the original sample of 23,994 participants, the author removed 918 participants that were likely duplicates, 1,455 participants with too many straightlining responses, and 688 participants with too many missing responses, leaving $n = \text{20,933}$ participants in the provided data set. Unfortunately, the raw data is not available at the above link. The minimal preprocessing that we applied to the provided data set is described in Appendix~\ref{app:application}.

In general, it seems likely that survey fatigue is present in a sample of responses to an online questionnaire with a lengthy battery of~300 items.
For instance, \citet{bowling2021length} estimate that if one wishes that at least~90\% of participants do not respond carelessly to more than~20\% of all items, one should not include more than 176 items in an online questionnaire.
By applying CODERS---which is specifically designed to identify partially careless responding through a change in response behavior---in addition to the traditional screeners applied by \citet{johnson2005}, new and complimentary insights into the incidence and nature of (partially) careless responding may emerge.

\subsection*{Results}

Throughout the discussion of the results, it is important to keep in mind that \citet{johnson2005} already removed straightliners from the publicly available data set. To maintain consistency in reporting, any percentages of participants refer to a baseline sample size of $N = \text{22,448}$ (i.e., including straightliners). For instance, the 1,455 straightliners found by \citet{johnson2005} amount to 6.5\% of participants.

\subsubsection*{General Overview}

\begin{figure}[!t]
\caption{\textit{Estimated onset of careless responding for participants flagged by CODERS in the data of \citet{johnson2005}.}}
\centering
\includegraphics[width = \textwidth]{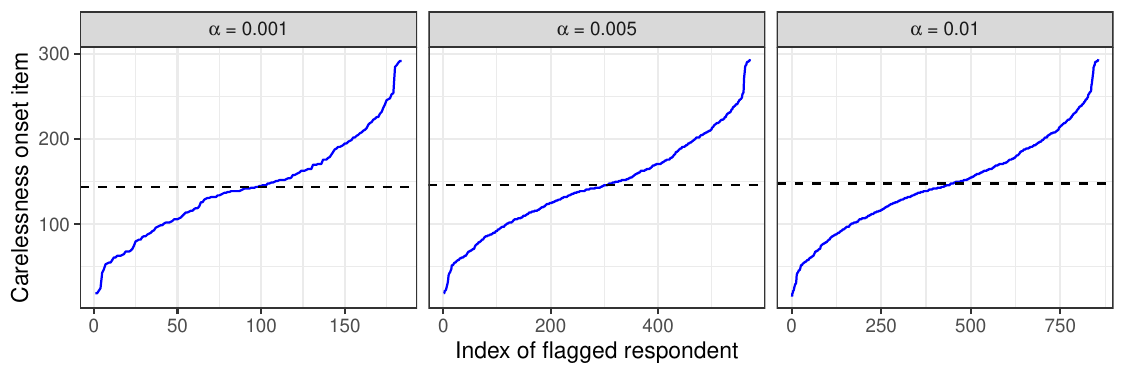}
\caption*{\textit{Note.} 
Results for different significance levels $\alpha \in\{0.01, 0.05, 0.1\}$ are presented in separate panels.
The $x$-axis shows the indices of the participants flagged as partially carelessness by CODERS. Here, the participants are sorted in ascending order of the estimated carelessness onset item, which is in turn depicted on the~$y$-axis. The dashed horizontal line represents the average location of carelessness onset across the flagged participants.}
\label{fig:application-location}
\end{figure}

For obtaining the autoencoder reconstruction errors in CODERS, we set the number of nodes in the bottleneck layer to~30, i.e., equal the number of facets in the Big Five model. 
Table~\ref{tab:application-CODERS} summarizes the results of CODERS. 
Depending on the significance level~$\alpha$, CODERS flags between 181~(0.8\%) participants for $\alpha = 0.001$ and 856~(3.8\%) participants for $\alpha = 0.01$ as partially careless.
On average, the location of the carelessness onset item for flagged participants is estimated around the~145th item, meaning that carelessness onsets on average after nearly half of the~300 items. 
The earliest onset occurs at item~15, whereas the latest onset is at item~294. 
Figure~\ref{fig:application-location} provides a more detailed visual overview of the detected onset items for the flagged participants at different significance levels.

\begin{table}[!b]
\caption{\textit{Results of CODERS for the data of \citet{johnson2005}.}}
\centering
\begin{tabular}{c l@{}r@{ }l c c c}
  \hline\noalign{\smallskip}
  & & & & \multicolumn{3}{c}{Detected onset} \\
	\noalign{\smallskip}\cline{5-7}\noalign{\smallskip}
	Significance level & \multicolumn{3}{c}{Number of flagged respondents} & 
	Minimum & Average & Maximum \\
	\noalign{\smallskip}\hline\noalign{\smallskip}
	$\alpha = 0.001$ & \phantom{Number o} & 181 & (0.8\%) & 19 & 143.7 & 292  \\
	$\alpha = 0.005$ &                    & 571 & (2.5\%) & 19 & 145.8 & 294 \\
	$\alpha = 0.01\phantom{0}$ &          & 856 & (3.8\%) & 15 & 147.9 & 294 \\
	\noalign{\smallskip}\hline\noalign{\smallskip}
\end{tabular}
\caption*{\textit{Note.} The methods are applied to the responses of $n = \text{20,993}$ participants to $p = 300$ items of a Big Five personality measurement. Among others, participants with too many straightlining responses have already been removed from the data provided by \citeauthor{johnson2005}. For consistent reporting, percentages refer to a baseline sample size of $N = \text{22,448}$, which includes straightliners.}
\label{tab:application-CODERS}
\end{table}

\begin{table}[!t]
\caption{\textit{Complementary findings of CODERS and traditional methods for the data of \citet{johnson2005}.}}
\centering
\begin{tabular}{l l@{}r@{ }l c c c}
  \hline\noalign{\smallskip}
  & & & & \multicolumn{3}{c}{Also flagged by CODERS} \\
	\noalign{\smallskip}\cline{5-7}\noalign{\smallskip}
	Method & \multicolumn{3}{c}{Number of flagged respondents} & 
	$\alpha = 0.001$ & $\alpha = 0.005$ & $\alpha = 0.01$ \\
	\noalign{\smallskip}\hline\noalign{\smallskip}
	Reliability & \phantom{Number o} & 140 & (0.6\%) & 2 & \phantom{1}8 & 10  \\
	Antonym     &                    & 284 & (1.3\%) & 9 &           18 & 24 \\
	Both        &                    &  11 &         & 1 & \phantom{1}1 & \phantom{2}1 \\
	\noalign{\smallskip}\hline\noalign{\smallskip}
\end{tabular}
\caption*{\textit{Note.} The methods are applied to the responses of $n = \text{20,993}$ participants to $p = 300$ items of a Big Five personality measurement. Among others, participants with too many straightlining responses have already been removed from the data provided by \citeauthor{johnson2005}. For consistent reporting, percentages refer to a baseline sample size of $N = \text{22,448}$, which includes straightliners. Too small percentages are omitted for better readability.}
\label{tab:application-agreement}
\end{table}

We now turn to analyzing the agreement between CODERS and the two traditional methods applied by \citet{johnson2005}, personal reliability and psychometric antonym. 
Following \citeauthor{johnson2005}, we reverse psychometric antonym scores so that low scores are associated with carelessness in both methods, and we flag respondents as careless whose personal reliability and psychometric antonym scores are lower than~$0.3$ and~$-0.03$, respectively.
The results are shown in Table~\ref{tab:application-agreement}. 
Personal reliability flags 140 participants (0.6\%) as careless, while psychometric antonym flags~284 respondents (1.3\%).\footnote{We use the \proglang{R} package \pkg{careless} \citep{careless} to compute personal reliability and psychometric antonym scores. The use of different software implementations (with potentially different default values) may explain differences with the numbers reported by \citet{johnson2005}.} 
CODERS, personal reliability, and psychometric antonyms generally show little agreement regarding the detected careless respondents. 
Personal reliability and psychometric antonym scores correlate by a low value of~0.314, yielding only 11 participants that are flagged by both methods. 
Depending on the significance level~$\alpha$, CODERS and personal reliability have between~2 and~10 flagged participants in common, while the number of jointly flagged respondents for CODERS and psychometric antonym ranges between~9 and~24. 
All three methods share a single flagged participant.

It is not surprising that there is little consensus of CODERS with the two traditional methods, as they are designed for different tasks. 
The traditional methods aim at identifying which respondents are careless, whereas CODERS aims to detect the onset of careless responding in partially careless respondents. 
Hence, this disagreement indicates that (i) there may be respondents who respond carelessly throughout the survey, which CODERS cannot identify due to the absence of a changepoint (but which are found by the traditional methods), and (ii) partial carelessness may be present in the data, for which traditional methods may have low detection power (but which are found by CODERS). 
Moreover, CODERS uncovers novel insights on careless responding behavior, which we highlight with a selection of respondents below.

\subsubsection*{Further Insights on Individual Cases}

\begin{figure}[!t]
\caption{\textit{Scores and changepoint item detected by CODERS in selected participants from the data of \citet{johnson2005}.}}
\centering
\begin{subfigure}[t]{0.49\textwidth}
\includegraphics[width = \textwidth]{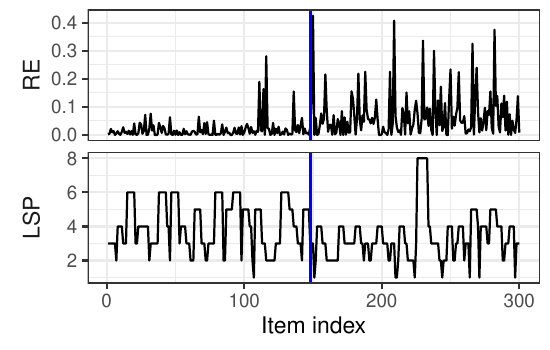}
\caption{Presumably inconsistent. Onset estimated by CODERS at item~148.}
\label{fig:appl-inconsistent-middle}
\end{subfigure}
\begin{subfigure}[t]{0.49\textwidth}
\includegraphics[width = \textwidth]{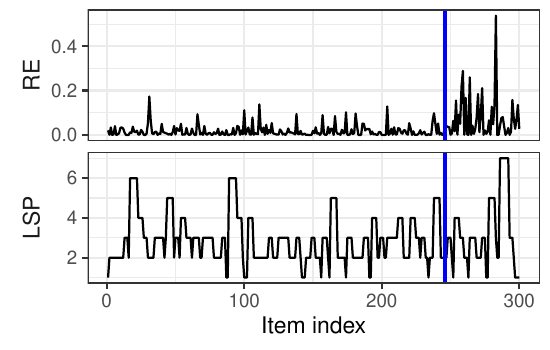}
\caption{Presumably inconsistent. Onset estimated by CODERS at item~246.}
\label{fig:appl-inconsistent-end}
\end{subfigure}
\\
\smallskip
\hfill
\begin{subfigure}[t]{0.49\textwidth}
\includegraphics[width = \textwidth]{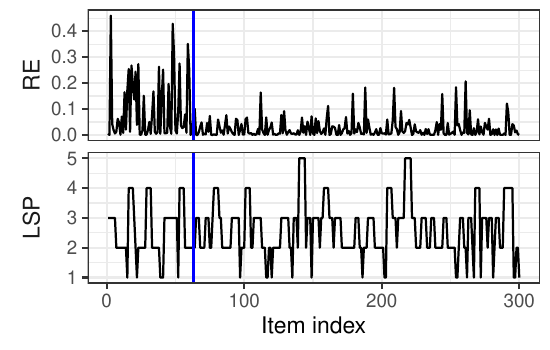}
\caption{Presumably inconsistent. Changepoint detected by CODERS at item~63.}
\label{fig:appl-inconsistent-begin}
\end{subfigure}
\begin{subfigure}[t]{0.49\textwidth}
\includegraphics[width = \textwidth]{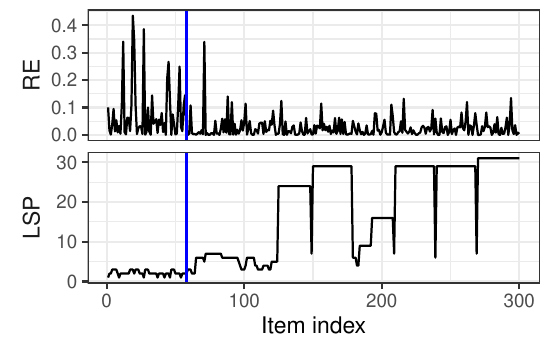}
\caption{Presumably inconsistent first, then invariable. Changepoint detected by CODERS at item~58.}
\label{fig:appl-inconsistent-invariable}
\end{subfigure}
\\
\smallskip
\hfill
\begin{subfigure}[t]{0.49\textwidth}
\includegraphics[width = \textwidth]{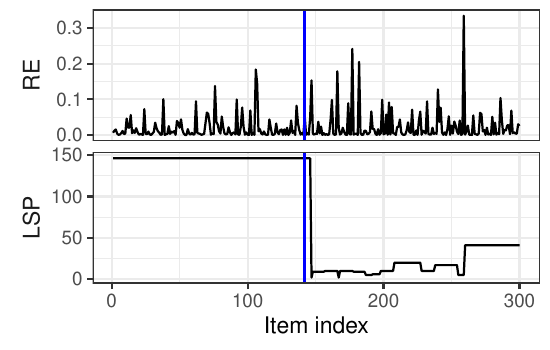}
\caption{Presumably invariable. Changepoint detected by CODERS at item~142.}
\label{fig:appl-invariable-throughout}
\end{subfigure}
\begin{subfigure}[t]{0.49\textwidth}
\includegraphics[width = \textwidth]{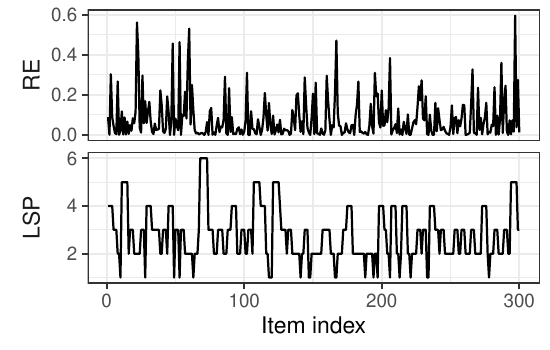}
\caption{Presumably inconsistent throughout the survey (no changepoint item detected by CODERS).}
\label{fig:appl-inconsistent-throughout}
\end{subfigure}
\caption*{\textit{Note.} In each subplot, autoencoder reconstruction errors (RE) and LongStringPattern (LSP) sequences are shown in separate panels. The solid vertical lines indicate the identified changepoint items. Summary statistics for the corresponding participants are provided in Table~\ref{tab:appl-dimavg}.}
\label{fig:appl-dimplots}
\end{figure}

To gain insights into the behavior of selected respondents, we analyze their per-item reconstruction errors and LSP sequences, which are shown in Figure~\ref{fig:appl-dimplots}. 
Table~\ref{tab:appl-dimavg} contains additional summary statistics for these measurements before and after the estimated changepoint item (if such an item is found by CODERS). 
While these examples are naturally selective, similar plots for all participants flagged by CODERS or the two traditional methods can be found in 
our online repository at \url{https://github.com/mwelz/carelessonset_plots}.

\begin{table}[!t]
\caption{\textit{Average scores before and after the changepoint item detected by CODERS in selected participants from the data of \citet{johnson2005}.}}
\begin{subtable}[t]{0.49\textwidth}
\centering
\begin{tabular}{l c c}
\hline\noalign{\smallskip}
    & Before        & After \\
\noalign{\smallskip}\hline\noalign{\smallskip}
RE  & 0.019 (0.035) & 0.069 (0.082) \\
LSP & 4.122 (1.428) & 3.373 (1.464) \\
\noalign{\smallskip}\hline\noalign{\smallskip}
\end{tabular}
\caption{Presumably inconsistent. Onset estimated by CODERS at item~148.}
\label{tab:appl-inconsistent-middle}
\end{subtable}
\begin{subtable}[t]{0.49\textwidth}
\centering
\begin{tabular}{l c c}
\hline\noalign{\smallskip}
    & Before        & After \\
\noalign{\smallskip}\hline\noalign{\smallskip}
RE  & 0.017 (0.026) & 0.070 (0.096) \\
LSP & 2.886 (1.189) & 3.236 (1.815) \\
\noalign{\smallskip}\hline\noalign{\smallskip}
\end{tabular}
\caption{Presumably inconsistent. Onset estimated by CODERS at item~246.}
\label{tab:appl-inconsistent-end}
\end{subtable}
\\
\medskip
\begin{subtable}[t]{0.49\textwidth}
\centering
\begin{tabular}{l c c}
\hline\noalign{\smallskip}
    & Before        & After \\
\noalign{\smallskip}\hline\noalign{\smallskip}
RE  & 0.095 (0.119) & 0.024 (0.036) \\
LSP & 2.565 (0.917) & 2.668 (0.925) \\
\noalign{\smallskip}\hline\noalign{\smallskip}
\end{tabular}
\caption{Presumably inconsistent. Changepoint detected by CODERS at item~63.}
\label{tab:appl-inconsistent-begin}
\end{subtable}
\begin{subtable}[t]{0.49\textwidth}
\centering
\begin{tabular}{l c c}
\hline\noalign{\smallskip}
    & Before        & After \\
\noalign{\smallskip}\hline\noalign{\smallskip}
RE  & 0.078 (0.103) & 0.025 (0.035) \\
LSP & 2.158 (0.621) & 19.708 (11.027) \\
\noalign{\smallskip}\hline\noalign{\smallskip}
\end{tabular}
\caption{Presumably inconsistent first, then invariable. Changepoint detected by CODERS at item~58.}
\label{tab:appl-inconsistent-invariable}
\end{subtable}
\begin{subtable}[t]{0.49\textwidth}
\centering
\begin{tabular}{l c c}
\hline\noalign{\smallskip}
    & Before        & After \\
\noalign{\smallskip}\hline\noalign{\smallskip}
RE  & 0.018 (0.029) & 0.024 (0.046) \\
LSP & 146.000 (0.000) & 23.597 (25.878) \\
\noalign{\smallskip}\hline\noalign{\smallskip}
\end{tabular}
\caption{Presumably invariable. Changepoint detected by CODERS at item~142.}
\label{tab:appl-invariable-throughout}
\end{subtable}
\begin{subtable}[t]{0.49\textwidth}
\centering
\begin{tabular}{l c}
\hline\noalign{\smallskip}
    & Overall \\
\noalign{\smallskip}\hline\noalign{\smallskip}
RE  & 0.079 (0.105) \\
LSP & 2.813 (1.115) \\
\noalign{\smallskip}\hline\noalign{\smallskip}
\end{tabular}
\caption{Presumably inconsistent throughout the survey (no changepoint item detected by CODERS).}
\label{tab:appl-inconsistent-throughout}
\end{subtable}
\caption*{\textit{Note.} Standard deviations are given in parentheses. The subtables correspond to the selected participants from Figure~\ref{fig:appl-dimplots}. Positioning of subtables is consistent with positioning of the subfigures in Figure~\ref{fig:appl-dimplots}, e.g., Table~\ref{tab:appl-inconsistent-middle} refers to the same participant as Figure~\ref{fig:appl-inconsistent-middle}.}
\label{tab:appl-dimavg}
\end{table}

First, we focus on the two participants in Figures~\ref{fig:appl-inconsistent-middle}--\ref{fig:appl-inconsistent-end} and Tables~\ref{tab:appl-inconsistent-middle}--\ref{tab:appl-inconsistent-end}. 
CODERS identifies carelessness onset at item~148 and item~246, respectively. 
We observe that the average reconstruction error, as well as the variability in the reconstruction errors, is much larger before the estimated onset item than afterward. 
Since the pre-onset responses could (generally) be reconstructed well while post-onset responses could not, there is evidence that these participants succumbed to survey fatigue and eventually engaged in inconsistent careless responding.

These two cases are representative for the majority of respondents flagged by CODERS, likely starting attentively but starting to respond carelessly at some point. 
In addition, more respondents flagged by CODERS seem to exhibit inconsistency than invariability, as \citeauthor{johnson2005} already removed straightliners from the publicly available data set. 
However, CODERS also identifies participants with remarkably different behavior.

The participant in Figure~\ref{fig:appl-inconsistent-begin} and Table~\ref{tab:appl-inconsistent-begin} shows similar behavior as the first two examples, albeit in reverse. 
CODERS identifies a changepoint at item~63, with reconstruction errors being large in terms of magnitude and variability before the changepoint while being small afterward. 
We conclude that this participant may have started the survey as an inconsistent careless respondent, but then shifted toward attentive responding.

For the participant in Figure~\ref{fig:appl-inconsistent-invariable} and Table~\ref{tab:appl-inconsistent-invariable}, CODERS identifies a changepoint at item~58. 
Both the reconstruction errors and the LSP sequence clearly change around this item, with even more excessive LSP values occurring in later items (maximum value of~31).
In fact, only four out of the $n=20,994$ participants in the sample yield more extreme LSP values.
Furthermore, the reconstruction errors before the changepoint are of comparable magnitude and variability to those of the previous examples of inconsistent responding.
Hence, there is strong evidence that this participant initially responded inconsistently, but switched to invariable responding characterized by (increasingly) lengthy sequences of fixed patterns as the questionnaire progressed.

Next, we discuss the example in Figure~\ref{fig:appl-invariable-throughout} and Table~\ref{tab:appl-invariable-throughout}, for whom CODERS detects a changepoint at item~142. 
It turns out that this respondent repeated the pattern \mbox{5-5-3-3-3} in the first half of the survey.
Even after the changepoint, the LSP values are among the highest in the sample, with the participant switching between the similar patterns \mbox{1-1-3-3-3}, \mbox{5-1-3-3-3}, and \mbox{1-5-3-3-3} for the remainder of the questionnaire. 
Although there is a clear change in behavior, as found by CODERS, this participant seems to respond invariably throughout the survey.

Finally, the example in Figure~\ref{fig:appl-inconsistent-throughout} and Table~\ref{tab:appl-inconsistent-throughout} corresponds to a participant flagged by psychometric antonym but not CODERS.  
A visual inspection confirms that there is indeed no clear changepoint in either the reconstruction errors or the LSP sequence. 
However, the reconstruction errors are similar in magnitude and variability to those found in inconsistent response periods from other examples. 
To investigate this further, we computed average reconstruction errors across all $p=300$ items for the $n=\text{20,993}$ participants. 
Only~6 participants have a higher average reconstruction error (cf.~Figure~\ref{fig:johnson2005density} in Appendix~\ref{app:application}), providing overwhelming evidence that this participant may be an inconsistent careless respondent throughout the whole questionnaire.

In summary, we reiterate that the first two examples are representative for the majority of participants flagged by CODERS, in line with the assumption that respondents start attentive and possibly become careless. 
Although other cases indicate that this assumption is too restrictive in practice, CODERS---which processes item-level indicators of inconsistency and invariability for changepoints---successfully uncovered relevant changes in response behavior. 
In light of these findings, we proceed to discuss limitations and potential extensions.


\section*{Limitations and Extensions}

We start by reviewing general limitations of CODERS together with (possible) remedies, before discussing the potential incorporation of response times into CODERS.

\subsection*{Limitations}\label{sec:limitations}

First, some participants may respond carelessly throughout all survey items (cf.~Figure~\ref{fig:appl-inconsistent-throughout} and Table~\ref{tab:appl-inconsistent-throughout}). Consequently, CODERS is unlikely to flag a changepoint due to an absence of a change in behavior. Nevertheless, the literature has established a plethora of methods for detecting careless respondents.
In this sense, CODERS---which is intended for detecting partial carelessness through a change in behavior---is complementary to those existing methods. If one suspects that respondents who have been careless throughout all items are present, we recommend to apply established detection methods following the guidelines of \citet{arthur2021} and \citet{ward2023}. Respondents that are flagged as careless by such methods, but for whom no changepoint was flagged by CODERS, have likely been careless throughout the survey.

Second, a detected changepoint does not necessarily indicate a change from attentive responding to careless responding. A participant may respond carelessly in the beginning of the survey and switch to responding attentively later on (cf.~Figure~\ref{fig:appl-inconsistent-begin} and Table~\ref{tab:appl-inconsistent-begin}), or they may switch between careless response behaviors through the course of the survey (cf.~Figures~\ref{fig:appl-inconsistent-invariable}--\ref{fig:appl-invariable-throughout} and Tables~\ref{tab:appl-inconsistent-invariable}--\ref{tab:appl-invariable-throughout}). For such cases, our empirical application demonstrates that CODERS---combined with a post-hoc analysis of the identified segments of responses---remains a valuable tool for gaining detailed insights into careless response behaviors.

Third, there may be  participants who alternate between periods of attentive and careless responding \citep{berry1992,clark2003,meade2012}. 
Our simulations indicate that CODERS remains conservative in this case and rarely flags one of the changepoints (see Appendix~\ref{app:simulations}).
CODERS should therefore be viewed as a promising first step towards detecting periods of careless responding, with an extension for the detection of multiple changepoints being an area of further research.

Finally, the changepoint detection procedure of \citet{shao2010} employed by CODERS is based on asymptotic arguments when the number of items is large. 
Hence, the number of items in the questionnaire is the main determinant for the statistical power of CODERS. For the detection of inconsistent careless responding, the number of participants plays a secondary role regarding power: the more attentive responses the autoencoder can learn from, the better the reconstruction errors can distinguish between attentive and careless responding  (cf.~Appendix~\ref{app:simulations}). The LSP sequence of a given participant, on the other hand, depends only on their own responses, meaning that the number of observations should not play a role concerning power in detecting invariable careless responding.
CODERS is thus primarily designed for relatively lengthy questionnaires, and our simulations suggest that there is a gradual drop in statistical power for detecting inconsistency as survey length decreases (see Appendix~\ref{app:simulations}).
However, partial carelessness due to survey fatigue may be less prevalent in short questionnaires \citep[cf.][]{bowling2021length}. 
A detailed study of partially careless responding in shorter surveys is therefore beyond the scope of this paper.

\subsection*{Response Times Revisited}

In addition to inconsistency and invariability, fast responding is considered the third major manifestation of carelessness \citep[e.g.,][]{ward2023}. In (online) surveys, response time is typically measured by the total time a participant spent on the questionnaire, time spent on each questionnaire page, or (less common) time spent on each questionnaire item.

If per-item response times are available, they can easily be incorporated into CODERS as an additional series to be used in the changepoint detection. If only per-page response times are available and pages are relatively short, an additional series can be obtained by assigning to each response the average page time, i.e., the time the participant spent on the page on which the item is located divided by the number of items on that page \citep[which is inspired by the page time index of][]{bowling2023time}. 
If a participant at some point in the survey, for instance, starts responding without reading the items, we expect a changepoint in response time towards considerably faster responses \citep[cf.][]{bowling2023time,bowling2016,meade2012,huang2012}.

Nevertheless, it is unclear if a changepoint in response times is indeed indicative of careless responding in an empirical setting. 
This may depend on the survey design: if items in one part of the survey are sufficiently longer than in other parts, a changepoint in response times might occur naturally for attentive respondents. 
On the other hand, if all items are of similar length, a changepoint to (much) shorter response times may indicate carelessness. For instance, there exist instruments whose individual items comprise of single words, such as the Big Five unipolar markers  \citep{goldberg1992} with items like \textit{``timid"} or \textit{``envious"}. In such measurements, one would not expect large differences in response times across items.
For such survey designs with similar item length, incorporating response times into CODERS may prove fruitful, but a thorough validation is beyond the scope of this paper and therefore left for future research.


\section*{Discussion}

We introduced a novel method, CODERS, designed to identify the onset of careless responding (or absence thereof) for each participant in a survey. 
It is motivated by literature stressing that survey length contributes to careless responding, in the sense that lengthy surveys---which are common in the behavioral sciences---may result in a large proportion of survey participants being partially careless, for instance due to fatigue or boredom. 
CODERS is highly flexible and can combine multiple indicators for the potential onset of carelessness, such as autoencoder reconstruction errors (RE) as an indicator of inconsistency, and the proposed LongStringPattern (LSP) sequence as an indicator of invariability. 
Simulation experiments indicate excellent performance of CODERS in identifying the onset of careless responding, and highlight the importance of incorporating multiple indicators that capture different manifestations of carelessness.
Re-analyzing data from a seminal study on careless responding, we demonstrate how CODERS reveals novel insights on partial carelessness. In particular when combined with a post-hoc analysis of the identified response segments, CODERS is able to uncover various types of changes in (careless) response behavior.

While CODERS is a promising step towards detecting periods of careless responding rather than identifying which respondents are careless, pertinent questions remain for future research. 
Most notably, extending or adapting CODERS with changepoint detection methods that allow for multiple changepoints, or that retain higher power in shorter surveys, may be fruitful directions for further research. 

CODERS naturally integrates with existing techniques that address careless responding. First, as illustrated in our empirical application, traditional post-hoc screeners and CODERS provide complementary insights. The former aim at identifying which respondents are careless, whereas CODERS specifically looks for changepoints in response behavior. It should be noted that some implementations allow traditional screeners to be computed on subsets of responses. For instance, the implementation of the intra-individual variance (IRV) \citep{dunn2018,marjanovic2015} in the \proglang{R} package \pkg{careless} \citep{careless} can be calculated for a given number of subsets of equal length. Although this approach is related to the detection of periods of careless responding, it analyzes predefined subsets of responses, while CODERS aims to identify the location of a changepoint.

Second, pre-administered detection items such as bogus, instructed, or self-report items \citep[e.g.][]{arthur2021, curran2016, ward2023} may provide further validation for the findings of CODERS. An advantage of detection items is that they can in principle detect any type of careless responding.
By themselves, detection items do not provide accurate estimates of the location of a change in response behavior, at least when following recommendations of \citet{meade2012} to include at most three detection items, spaced every 50–100 items. 
However, detection items can strengthen the findings of CODERS: if passed detection items (i.e., responses expected from an attentive participant) fall with the found attentive period while failed detection items fall within the found careless period, there is convergent evidence of such response behavior.

Third, various methods have been proposed in the literature to prevent careless responding in the first place, such as promising rewards for responding accurately \citep{gibson2020}, or threatening with punishment for careless responding  \citep{bowling2021length}. There is evidence that such measures can prevent or at least delay the onset of careless responding \citep{bowling2021length,gibson2020,huang2012}. However, preventive measures can be undesirable or even unethical in practice. For instance, \citet{arthur2021} and \citet{bowling2021length} advise against using threats if one wishes to maintain a positive long-term relationship with survey participants, while \citet{arthur2021} stress that promising rewards may cross ethical boundaries in some contexts. Notwithstanding, if preventative methods are appropriate, combining them with screening techniques such as CODERS may further enhance quality and reliability of the collected data for subsequent analysis.

Whereas we apply machine learning and nonparametric techniques in CODERS, other literature is focused on model-based approaches. 
For instance, \citet{ulitzsch2022notime} propose a mixture model based on item response theory that is designed to detect careless respondents, which is adapted in \citet{ulitzsch2022time} to incorporate response times. 
Other approaches that explicitly model carelessness as part of a mixture model are proposed in \citet{zhang2025}, \citet{steinmann2022}, \citet{vanlaar2022}, and \citet{arias2020}. One advantage of model-based approaches is that their estimates can be used to construct person-weights that can be used to downweight the adverse influence of potentially careless respondents in subsequent analyses, which is discussed in detail in \citet{ulitzsch2024, ulitzsch2022time,ulitzsch2022notime} and \citet{hong2019}. However, model-based approaches typically impose strong and potentially unverifiable assumptions on the properties of attentive and careless responding.

On a final note, \citet{arthur2021} envisage  \textit{``artificial intelligence and machine learning being used to detect careless responding and response distortion in real time in the foreseeable future."} 
Using machine learning techniques in the form of autoencoders and nonparametric changepoint detection, our proposed method CODERS takes a step towards that future.

\section*{Computational Details}
Throughout this paper, we have followed recent guidelines by \citet{alfons2024} on the application of machine learning methods for the detection of careless responding. All computations were performed with~\proglang{R} version~4.4.2 \citep{R} and \proglang{Python} version~3.10.12 \citep{python3}. Our implementation of CODERS in the \proglang{R} package
\pkg{carelessonset} \citep{R:carelessonset}
is available from \url{https://github.com/mwelz/carelessonset}.
\emph{Replication files will be made publicly available upon acceptance of this manuscript and a link will be included here.}

\section*{Acknowledgments}
We thank Donald Bergh, Dennis Fok, Patrick Groenen, Nick Koning, Robin Lumsdaine, Jakob Raymakers, and Kevin ten Haaf for valuable comments and feedback. This work was supported by a grant from the Dutch Research Council (NWO), research program Vidi (Project No. VI.Vidi.195.141).


\bibliography{bibliography}


\appendix

\newpage
\renewcommand\thefigure{\thesection\arabic{figure}}
\renewcommand\thetable{\thesection\arabic{table}}
\renewcommand{\theequation}{\thesection\arabic{equation}}
\renewcommand\thealgocf{\thesection\arabic{algocf}}
\renewcommand{\thefootnote}{\roman{footnote}}

\section{Setup and Assumptions}
\label{app:setup}

\setcounter{figure}{0}
\setcounter{table}{0}
\setcounter{equation}{0}
\setcounter{footnote}{0}

Throughout our paper, we consider the following setup. Let~$\mat{X}$ be an $n\times p$ data matrix holding the rating-scale responses of~$n$ survey participants to~$p$ items. 
We make the standard assumption that all observations (i.e., all participants whose responses the survey collects) are independently and identically distributed.
Moreover, it is a priori unknown if and when careless responding occurs in~$\mat{X}$.
We require the following assumptions. Importantly, these assumptions should not be understood as a conceptual model for careless responding, but rather as statistical prerequisites for CODERS to work best. In the immortal words of John von Neumann, ``[the truth] is much too complicated to allow anything but approximations.''
 
\begin{assumption}\label{ass:lowdim}
The responses in $\mat{X}$ admit an $n \times s$ lower-dimensional representation~$\Blowdim$, where $s \ll p$. The dimension $s$ is known and corresponds to the number of constructs that the survey measures. 

This assumption is very mild in survey data since surveys typically measure multiple constructs, with the number of constructs typically known to the survey designer. For instance, the $p=300$ item representation of the NEO-PI-R inventory \citep{goldberg1999} measures $s=30$ constructs.
\end{assumption}
 
\begin{assumption}\label{ass:reliable}
The survey that generated $\mat{X}$ is reliable in the sense that if all participants responded accurately, $\mat{X}$ would accurately measure all constructs. 

This assumption is again mild in survey data. There exist well-established and reliable survey measures for a large variety of constructs. For instance, the International Personality Item Pool \citep[IPIP;][]{goldberg1999} is a pool of more than~250 personality scales. It is in general recommended to use well-established measures to ensure a highly reliable measurement \citep[p.~150]{babbie2020}. The reliability of a measurement can be estimated in multiple ways, such as Cronbach's alpha, McDonald's omega and variations thereof, as well as the $H$~coefficient \citep[see, e.g.,][for a description of these and other reliability measures]{mcneish2018}. Hence, this assumption is satisfied when the survey data are collected with measures of high reliability.
\end{assumption}

\begin{assumption}\label{ass:randomized}
The survey design precludes responses in (fixed) recurring patterns, meaning that such patterns are indicative of invariable careless responding.

One way to satisfy this assumption is through randomization of the survey items, which is common practice in surveys in the behavioral sciences. If it is desirable to keep items of the same construct together to reduce the burden on respondents, one may randomize the order of items within each construct, and randomize the order of constructs (or groups of constructs, if related constructs should be presented together). The use of negatively keyed items may also be beneficial.
\end{assumption}
 
\begin{assumption}\label{ass:endurance}
Participants begin the survey as attentive respondents by providing accurate and truthful responses. As the survey progresses, some (possibly none or all) participants start responding carelessly and continue to do so for the remainder of the survey.

Under this assumption, a detected changepoint (see also the next assumption) naturally divides a participant's responses into an attentive segment before the changepoint and a careless segment from the changepoint onward. In practice, a changepoint may characterize different changes in response behavior. For instance, a participant may engage in careless responding in the beginning of the survey but later switch to attentive responding, or they may switch from one type of careless responding to another (e.g., from inconsistency to invariability). Via a post-hoc analysis of the identified response segments, CODERS remains useful for uncovering insights on such changes in response behavior, as highlighted in the empirical application with selected cases (cf.~Figures~\ref{fig:appl-inconsistent-begin}--\ref{fig:appl-invariable-throughout} and Tables~\ref{tab:appl-inconsistent-begin}--\ref{tab:appl-invariable-throughout} in the main text).

It is also conceivable that a respondent may switch repeatedly between attentive and careless responding. For instance, they may start attentively, then be distracted for a certain period of time, but switch back to attentive responding once the distraction goes away. CODERS in its current form is not designed to pick up such behavior, but an extension using multiple changepoint detection may be a fruitful venue for future research.
\end{assumption}

\begin{assumption}\label{ass:breaks}
The onset of partial carelessness is characterized by a changepoint in one or more indicators. We use autoencoder reconstruction errors (RE) and LongStringPattern (LSP) sequences as indicators. 

A detailed motivation for this assumption is provided in the section of the main text that introduces CODERS.
\end{assumption}


\clearpage

\section{Details on CODERS}
\label{app:methodology}

\setcounter{figure}{0}
\setcounter{table}{0}
\setcounter{equation}{0}

\subsection*{Motivating Novel Methodology}

Before we present a technical description of CODERS, we believe that it is instructive to first take a step back and motivate the application of novel techniques for the task of detecting partially careless responding rather than using existing post-hoc screeners such as personal reliability, psychometric antonym, and the longstring index \citep[see e.g.,][for a recent overview of traditional screeners]{alfons2024}. 

A primary motivation for devising new methodology is that most existing screeners compute a single carelessness score for each participant. 
While this can work very well for distinguishing careless from attentive respondents, having a singular score for every respondent does not provide much information beyond the attentive-careless classification. 
As outlined in the main text, recent literature suggests that partially careless responding might be a common occurrence.
As such, it might be more informative to know \emph{when} a given respondent starts responding carelessly (if at all) rather than a dichotomous attentive-careless distinction. 
For instance, this could be useful for knowing when the responses of a given respondent become unreliable so that one must not drop all of their responses in a primary analysis. 
In order to obtain a granular estimate of carelessness onset on the participant level, we require a carelessness score for \emph{every} item for \emph{every} respondent rather than a singular per-participant carelessness score. 
While some traditional screeners can be somewhat tweaked in this direction \citep[for instance, one could calculate separate intra-individual response variances for separate sets of items; see e.g.,][]{dunn2018,marjanovic2015},  accurate estimation of the onset of carelessness necessitates new methodology.

As such, CODERS is based on two different carelessness scores on the participant-item level to capture a wide variety of different types of careless responding. 
The first dimension, reconstruction errors (RE) obtained via an auto-associative neural network (autoencoder) \citep{kramer1992}, is intended to measure inconsistent careless responding, such as choosing response categories at random, on the participant-item level. 
The second dimension, a LongStringPattern (LSP) sequence, is a generalization of the famous longstring index \citep{johnson2005} that measures on the participant-item level if a given response is part of a content-independent recurring response sequence, such as $1-2-1-2-\dots-1-2$. 
If so, it allows one to trace back when the recurring sequence has started, thereby providing evidence for the onset of invariable carelessness. 
Autoencoders and LSP are described in detail in the following sections.

\subsection*{Details on Autoencoders}

\subsubsection*{Further Intuition}

Assigning a participant-item score that measures how inconsistently 
a participant has responded to the item in question arguably is a challenging task. Since different attentive participants may respond differently to a given item (between-participant heterogeneity), one needs to somehow understand how attentive responding manifests on the participant-level so that a possible switch to inconsistent responding can be detected. Otherwise, there would be the risk of confusing attentive responding for inconsistent responding, and vice versa. We argue that an auto-associative neural network (autoencoder) is useful for learning within-participant attentive response structures.

An autoencoder as used in CODERS essentially predicts the response of each respondent to each item. The underlying idea is that if the predicted response is notably different from the given response, the participant has given a response that is unexpected, judging from their previous response behavior. While an occasional unexpected response is not out of the ordinary, having multiple such unexpected responses occur consecutively may suggest that the participant has shifted from predictable responding to unpredictable responding, which might be viewed as indication of an onset of inconsistent responding. Since we assume that every participant has provided attentive responses at the beginning of the questionnaire (see Assumption~\ref{ass:endurance}), it should be conceptually possible to learn how attentive responding manifests for each participant so that somewhat accurate response predictions can be made up to carelessness onset. 

The autoencoder learns attentive response behavior by exploiting a key feature of questionnaires, namely that multiple items measure the same construct through psychometric scales. Since one usually knows how many scales are measured by the survey at hand, the information contained in a given participant's responses should be representable in much fewer dimensions than the total number of items (see Assumption~\ref{ass:lowdim}). 
Thus, the task of learning attentive response behavior can be seen as finding a good lower-dimensional representation for every participant's given responses. 

Formally, the autoencoder minimizes the participant-item response prediction error over all participants and items. However, in the process of making predictions, the autoencoder enforces to first find a lower-dimensional representation of each participant's given responses, and then to make a response prediction based on that lower-dimensional representation. In other words, the autoencoder first needs to compress~$p$ dimensions into~$s\ll p$ dimensions, and then decompress again that lower-dimensional representation to obtain $p$-dimensional predictions. The potential presence of inconsistent careless responses should not much affect the quality of the lower-dimensional representation---unless carelessness onsets very early---because inconsistent responding is characterized by its random-like nature, meaning that such responses are not informative for learning the lower-dimensional representation.

We stress that response predictions based on a lower-dimensional representation could also be obtained through principal component analysis (PCA). In fact, autoencoders turn out to be a generalization of PCA that, unlike PCA, is not based on linearity assumptions \citep{kramer1991}. Since we wish to refrain from possibly restrictive assumptions, we choose autoencoders over PCA to construct lower-dimensional representations of the given responses. 
We are now ready to dive into the formal definition of the autoencoder.

\subsubsection*{Network Architecture and Estimation}

Denote by~$x_{ij}$ the response of the~$i$-th participant to the~$j$-th survey item. Collect the responses of the~$i$-th participant to all~$p$ items in a $p$-dimensional vector~$\obs{x}_i = (x_{i1}, x_{i2}, \dots, x_{ip})^\top$ so that the~$n\times p$ data matrix~$\mat{X}$ is given by~$\mat{X} = (\obs{x}_1, \obs{x}_2, \dots,\obs{x}_n)^\top$.
 
Introduced by \citet{kramer1992}, an autoencoder is a network consisting of an uneven number of~$M$ layers, in which the first layer (\emph{input layer}) holds the input data~$\obs{x}_i$ and the last layer (\emph{output layer}) holds the reconstructed input data~$\widehat{\obs{x}}_i$. 
Denoting the number of nodes in the~$\ell$-th layer by $N^{(\ell)}$, we have that $N^{(1)} = N^{(M)} = p$. Moreover, the central layer (\emph{bottleneck layer}) consists of $s<p$ nodes to prevent that the input is simply passed through the network, and the network architecture is symmetric around the bottleneck layer.

Each node in a given layer stores so-called \emph{activations}, which are obtained by transforming a linear combination of the activations from the previous layer. This transformation is defined by an \emph{activation function} $g_\ell (\cdot)$. Formally, for participant $i=1,\dots,n$, the activation $a_{ij}^{(\ell)}$ of the $j$-th node, $j=1,\dots, N^{(\ell)},$ of the $\ell$-th layer, $\ell=2,\dots, M$, is given by
\begin{align*}
a_{ij}^{(\ell)} &= g_\ell \left(  z_{ij}^{(\ell)} \right), 
\\
z_{ij}^{(\ell)} &=
\sum_{k=1}^{N^{(\ell-1)}} \omega_{jk}^{(\ell)} a_{ik}^{(\ell-1)} + b_j^{(\ell)}.
\end{align*}
The $\omega_{jk}^{(\ell)}$ and $b_j^{(\ell)}$ are unknown \textit{weights} and  \textit{intercept} terms, respectively. 
Observe that in the first layer, we have $a_{ij}^{(1)} = x_{ij}$. The activations of the last layer, $a_{ij}^{(M)}$, hold the network's output, namely the autoencoder's reconstructions~$\widehat{x}_{ij}$ of responses~$x_{ij}$. 

We jointly refer to the weights and intercept terms as the network's \textit{parameters}, which need to be estimated, and we collect them in a vector~$\Btheta$. 
To emphasize the dependence on the parameter vector~$\Btheta$, we define for a fixed~$\Btheta$ a prediction function~$\Bf_{\Btheta}(\cdot)$ of the network, which corresponds to the activations in the last layer:
\[
 	\Bf_{\Btheta}  (\obs{x}_i) =
 	\Big( f_{1,\Btheta}(\obs{x}_i), f_{2,\Btheta}(\obs{x}_i), \dots, f_{p,\Btheta}(\obs{x}_i)  \Big)^\top =
     \left( a_{i1}^{(M)}, a_{i2}^{(M)}, \dots, a_{ip}^{(M)} \right)^\top,
\]
for participants $i = 1,\dots,n$. The reconstruction of the input vector~$\obs{x}_i$ is then given by~$\widehat{\obs{x}}_i~=~\Bf_{\Btheta} (\obs{x}_i)$.
Due to the fact that each node is a function of nodes from a previous layer, the nodes in adjacent layers are often visualized by connecting edges. 
Figure~\ref{fig:nn} provides such a schematic overview of 
a general autoencoder neural network.

\begin{figure}[!t]
 \caption{\textit{Schematic overview of an autoencoder neural network with~$M$ layers.}}

 	\centering
 	\begin{tikzpicture}[shorten >=1pt]
 		\tikzstyle{unit}=[draw,shape=circle,minimum size=1.15cm]
 		\tikzstyle{hidden}=[draw,shape=circle,minimum size=1.15cm]
 
 		\node[unit, fill=EURwarmgray](l11) at (-0.5,3.5){$x_{i1}$};
 		\node[unit, fill=EURwarmgray](l12) at (-0.5,2){$x_{i2}$};
 		\node at (-0.5,1){\vdots};
 		\node[unit, fill=EURwarmgray](l13) at (-0.5,-0.17){$x_{ip}$};
 
 		\node[unit, fill=EURwarmgray](l21) at (2.5,5){$a_{i1}^{(2)}$};
 		\node[unit, fill=EURwarmgray](l22) at (2.5,3){$a_{i2}^{(2)}$};
 		\node[unit, fill=EURwarmgray](l23) at (2.5,1){$a_{i3}^{(2)}$};
 		\node at (2.5,0){\vdots};
 		\node[unit, fill=EURwarmgray](l24) at (2.5,-1.6){$a_{i,N^{(2)}}^{(2)}$};
 
 		\node(h31) at (4.5,0){};
 		\node(h32) at (4.5,2){};
 		\node(h33) at (4.5,4){};
 
 		\node(d3) at (5.5,0){$\ldots$};
 		\node(d2) at (5.5,2){$\ldots$};
 		\node(d1) at (5.5,4){$\ldots$};
 
 		\node(h41) at (6.5,0){};
 		\node(h42) at (6.5,2){};
 		\node(h43) at (6.5,4){};
 
 		\node[unit, fill=EURwarmgray](lX1) at (8.5,5){$a_{i1}^{(M-1)}$};
 		\node[unit, fill=EURwarmgray](lX2) at (8.5,3){$a_{i2}^{(M-1)}$};
 		\node[unit, fill=EURwarmgray](lX3) at (8.5,1){$a_{i3}^{(M-1)}$};
 		\node at (8.5,-0.1){\vdots};
 		\node[unit, fill=EURwarmgray](lX4) at (8.5,-1.6){$a_{i,N^{(M-1)}}^{(M-1)}$};
 
 		\node[unit, fill=EURwarmgray](out1) at (11.5,4.2){$f_{1,\Btheta}(\obs{x}_i)$};
 		\node[unit, fill=EURwarmgray](out2) at (11.5,2.2){$f_{2,\Btheta}(\obs{x}_i)$};
 		\node at (11.5,1){\vdots};
 		\node[unit, fill=EURwarmgray](out3) at (11.5,-0.5){$f_{p,\Btheta}(\obs{x}_i)$};
 
 		\draw[->] (l11) -- (l21);
 		\draw[->] (l11) -- (l22);
 		\draw[->] (l11) -- (l23);
 		\draw[->] (l11) -- (l24);
 
 		\draw[->] (l12) -- (l21);
 		\draw[->] (l12) -- (l22);
 		\draw[->] (l12) -- (l23);
 		\draw[->] (l12) -- (l24);
 
 		\draw[->] (l13) -- (l21);
 		\draw[->] (l13) -- (l22);
 		\draw[->] (l13) -- (l23);
 		\draw[->] (l13) -- (l24);
 
 		\draw[->,path fading=east] (l21) -- (h31);
 		\draw[->,path fading=east] (l21) -- (h32);
 		\draw[->,path fading=east] (l21) -- (h33);
 
 		\draw[->,path fading=east] (l22) -- (h31);
 		\draw[->,path fading=east] (l22) -- (h32);
 		\draw[->,path fading=east] (l22) -- (h33);
 
 		\draw[->,path fading=east] (l23) -- (h31);
 		\draw[->,path fading=east] (l23) -- (h32);
 		\draw[->,path fading=east] (l23) -- (h33);
 
 		\draw[->,path fading=east] (l24) -- (h31);
 		\draw[->,path fading=east] (l24) -- (h32);
 		\draw[->,path fading=east] (l24) -- (h33);
 
 		\draw[->,path fading=west] (h41) -- (lX1);
 		\draw[->,path fading=west] (h41) -- (lX2);
 		\draw[->,path fading=west] (h41) -- (lX3);
 		\draw[->,path fading=west] (h41) -- (lX4);
 
 		\draw[->,path fading=west] (h42) -- (lX1);
 		\draw[->,path fading=west] (h42) -- (lX2);
 		\draw[->,path fading=west] (h42) -- (lX3);
 		\draw[->,path fading=west] (h42) -- (lX4);
 
 		\draw[->,path fading=west] (h43) -- (lX1);
 		\draw[->,path fading=west] (h43) -- (lX2);
 		\draw[->,path fading=west] (h43) -- (lX3);
 		\draw[->,path fading=west] (h43) -- (lX4);
 
 		\draw[->] (lX1) -- (out1);
 		\draw[->] (lX2) -- (out1);
 		\draw[->] (lX3) -- (out1);
 		\draw[->] (lX4) -- (out1);
 
 		\draw[->] (lX1) -- (out2);
 		\draw[->] (lX2) -- (out2);
 		\draw[->] (lX3) -- (out2);
 		\draw[->] (lX4) -- (out2);
 
 		\draw[->] (lX1) -- (out3);
 		\draw[->] (lX2) -- (out3);
 		\draw[->] (lX3) -- (out3);
 		\draw[->] (lX4) -- (out3);
 
 		\draw [decorate,decoration={brace,amplitude=6pt},xshift=0pt,yshift=0pt] (-1,4.2) -- (0,4.2) node [black,midway,yshift=+0.5cm]{Input};
 
 		\draw [decorate,decoration={brace,amplitude=6pt},xshift=0pt,yshift=0pt] (2,5.7) -- (3,5.7) node [black,midway,yshift=+0.5cm]{2nd layer};
 
 		\draw [decorate,decoration={brace,amplitude=6pt},xshift=0pt,yshift=0pt] (8,5.9) -- (9,5.9) node [black,midway,yshift=+0.5cm]{$(M-1)$-th layer};
 
     	 \draw [decorate,decoration={brace,amplitude=6pt},xshift=0pt,yshift=0pt] (11,5.2) -- (12,5.2) node [black,midway,yshift=+0.5cm]{Output};
 
 	\end{tikzpicture}
 \caption*{\textit{Note.} The first and last layer are referred to as input and output layer, respectively. For participants $i=1,\dots,n$, the input in the first layer are the input responses $\obs{x}_i = (x_{i1},\dots,x_{ip})^\top$, and the output in the last layer are the activations $f_{j,\Btheta}(\obs{x}_i) = a_{ij}^{(M)},\ j=1,\dots,p,$ that reconstruct the input responses for a parameter vector $\Btheta$ to be estimated. The $j$-th node in the $\ell$-th layer holds the activation $a_{ij}^{(\ell)}$, for $\ell = 2,\dots,M, j=1,\dots,N^{(\ell)}$. Besides the parameter vector $\Btheta$, the prediction function $\Bf_{\Btheta}(\cdot)$ depends implicitly on the prespecified choice of the activation functions $g_\ell(\cdot)$.}
 \label{fig:nn}
\end{figure}
 
To fit an autoencoder, we aim at minimizing its reconstruction error. For a prespecified loss function $\Ell (\cdot)$, we fit the neural network by finding the $\Btheta$ that yields the best average reconstruction error,
\begin{equation}\label{eq:loss-unpenalized}
 	\widehat{\Btheta} = \arg\min_{\Btheta}
 	\Bigg\{  \frac{1}{n}
 	\sum_{i=1}^n \sum_{j=1}^p
 	 \Ell \Big( x_{ij} - f_{j,\Btheta}(\obs{x}_i) \Big)
 	 \Bigg\}.
\end{equation}
The prediction function of the fitted network is subsequently given by~$\Bf_{\widehat{\Btheta}}(\cdot)$. For the loss function in~\eqref{eq:loss-unpenalized}, we choose the smooth and robust \textit{Pseudo-Huber} loss, which is defined for a fixed $\delta >0$ as
\begin{equation}\label{eq:huberloss}
 	\Ell(z)
 	=
 	\delta^2\left( \sqrt{1 + (z/\delta)^2} - 1 \right).
\end{equation}
This choice results in quadratic loss for small values of~$z$ and linear loss for large values of~$z$. Consequently, the Pseudo-Huber loss function avoids that large individual reconstruction errors strongly affect the fit, which would make it hard for the network to distinguish attentive responses from inconsistent careless responses, as the latter typically lead to large prediction errors. To solve the optimization problem in \eqref{eq:loss-unpenalized}, we use stochastic gradient descent. We refer to Chapter~8.5 in \citet{goodfellow2016} for details.

\subsubsection*{Modeling Choices}

Following \citet{kramer1992}, we specify $M=5$ layers in the autoencoder. 
We set the number of nodes in layers~2 and~4 (mapping an demapping layer, respectively) to $\lfloor 1.5\times p \rfloor$, as a relatively large number of nodes is expected to give the autoencoder flexibility to learn many different types of response behavior (attentive and careless).
Based on Assumption~\ref{ass:lowdim}, the number of nodes in the bottleneck layer equals the number of constructs measured in the questionnaire.
 
Concerning the activation functions, we again follow a recommendation in \citet{kramer1992} and propose to use nonlinear activation functions in the mapping and demapping layers, as well as a linear activation function in the bottleneck layer. 
Specifically, we propose to use the hyperbolic tangent activation in the (de-)mapping layers and the identity mapping in the bottleneck layer. 
Definitions are given in Table~\ref{tab:ANN-architecture}, which summarizes our proposed autoencoder architecture.
 
\begin{table}[!t]
\caption{\textit{Choices regarding the network architecture of the autoencoder.}}
\centering
\begin{tabular}{l c l}
\hline\noalign{\smallskip}
Layer & Number of nodes & Activation function \\
\noalign{\smallskip}\hline\noalign{\smallskip}
Mapping layer & $\lfloor 1.5 \times p \rfloor$ &
$\textsf{tanh}(x) = 2/(1 + \exp (-2x)) - 1$\smallskip \\
Bottleneck layer & $s$ &
$\textsf{identity}(x) = x$\smallskip \\
Demapping layer & $\lfloor 1.5 \times p \rfloor$ &
$\textsf{tanh}(x) = 2/(1 + \exp (-2x)) - 1$ \\
\noalign{\smallskip}\hline
\end{tabular}
\caption*{\textit{Note.} The constants~$p$ and~$s$ denote the number of items in the questionnaire and the number of constructs that the questionnaire measures, respectively. For a real number~$x$, $\lfloor x \rfloor$ denotes the largest integer smaller than or equal to~$x$.}
\label{tab:ANN-architecture}
\end{table}
 
For fitting the autoencoder, we use the stochastic gradient descent algorithm with a batch size of~10, a learning rate of~0.0001, and~100 epochs. In the pseudo-Huber loss~\eqref{eq:huberloss}, we set constant~$\delta = 1$.

\subsection*{LongStringPattern Algorithm}

Formal algorithms for computing the $l$-pattern and LSP sequences are provided in Algorithms~\ref{alg:longstringpattern} and~\ref{alg:adaptivelongstringpattern}, respectively.

\begin{algorithm}  
 	\caption{\texttt{$l$-pattern}($\{x_j\}_{j=1}^p, l$)}\label{alg:longstringpattern}
 	\KwIn{Series of observed rating-scale responses $\{x_j\}_{j=1}^p$, pattern length $l\in \N$ to consider (number of items that an individual pattern encompasses).
 	 For instance, the individual pattern $1-2-3$ encompasses $l=3$ items in the response sequence $1-2-3-1-2-3$. 
 	Note that the algorithm measures straightlining when setting $l=1$.}
 	\KwOut{Sequence of measures $\{l\textsf{-pattern}_j\}_{j=1}^p$}
 	\KwInp{$streak \gets 1,startindex \gets 1$}
 
 	\BlankLine
 	\tcc{fill in the first $l-1$ indices}
 	\For{$k\gets 1$ \KwTo $p-l-1$ \KwBy $1$}
 	{
 	 \eIf{$x[k] == x[k+l]$}{increment $streak$ by 1 \tcp*[l]{streak is ongoing}}
 	 {
 	   Assign value $streak$ to all elements in $\{l\textsf{-pattern}_j\}_{j=startindex}^k$
 	   \tcp*[l]{streak broke}
 	   $startindex \gets k + 1$\;
 	   streak $\gets 1$\;
 	 }
 	}
 
 	\BlankLine
     \tcc{fill in the last $l$ indices (there are 3 edge cases)}
     \uIf{$\texttt{streak} \geq l$}
     {
      \tcc{case 1: current streak is ongoing and we have seen it completely at least once, so assume it will last until the end and complete the current streak (which has not yet been broken)}
      Assign value $streak+l-1$ to all elements in $\{l\textsf{-pattern}_j\}_{j=startindex}^p$\;
     }
     \uElseIf{$1 <\texttt{streak} < l$}{
     \tcc{case 2: there is an ongoing streak, but we have no chance of seeing it completely (since streak < l), so we do not know how the streak looks like. In this case, fill in the current value of "streak" until the (p-l)-th item and fill in 1 for the last l items (because we do not know what the complete streak looks like)}
     Assign value $streak-1$ to all elements in $\{l\textsf{-pattern}_j\}_{j=startindex}^{p-l}$\;
     Assign value $1$ to all elements in $\{l\textsf{-pattern}_j\}_{j=p-l+1}^{p}$\;
     }
     \Else{
     \tcc{case 3: there is no ongoing streak, so fill in 1 for last $l$ indices}
     Assign value~1 to all elements in $\{l\textsf{-pattern}_j\}_{j=p-l+1}^p$\;
     }
\end{algorithm}

\begin{algorithm}[!t]
 	\caption{\texttt{LongStringPattern}($\{x_j\}_{j=1}^p, l_{max}$)}\label{alg:adaptivelongstringpattern}
 	\KwIn{Series of observed rating-scale responses $\{x_j\}_{j=1}^p$, maximum pattern length $l_{max}\in \N$ to consider}
 	\KwOut{Set of LongStringPattern measures $\{\textsf{LSP}_j\}_{j=1}^p$}
 
 	\BlankLine
 	\tcc{calculate $l$-pattern for fixed pattern length $l$}
 	\For{$l\gets 1$ \KwTo $l_{max}$ \KwBy $1$}
 	{
 	  $\{\textsf{LSP}^l_j\}_{j=1}^p \gets l\textsf{-pattern}(\{x_j\}_{j=1}^p, l)$\;
 	}
 
 	\BlankLine
 	\tcc{for each element $\{1,\dots,p\}$, report the largest $l$-pattern value}
 	Initialize $\{\textsf{LSP}_j\}_{j=1}^p$\;
 	\For{$j\gets 1$ \KwTo $p$ \KwBy $1$}
 	{
 	  $\textsf{LSP}_j \gets \max\big\{ \textsf{LSP}_j^1, \textsf{LSP}_j^2,\dots, \textsf{LSP}_j^l \big\}$\;
 	}
\end{algorithm}

\subsection*{Details on Changepoint Detection}

Let $\{\vect{Y}_j\}_{j=1}^p$ be a series of length $p$ consisting of $d$-dimensional random variables. 
The goal of the cumulative sum self-normalization test of \citet{shao2010} is to estimate the location of a possible changepoint~$k$ in the (multivariate) series, i.e., to divide the series into two subsets $\{\vect{Y}_j\}_{j=1}^{k-1}$ and $\{\vect{Y}_j\}_{j=k}^p$ in which a $d$-dimensional parameter of interest~$\Btheta$ differs.  
Here, we focus on the mean as the parameter~$\Btheta$. 
Define by $\widehat{\Btheta}_{a,b} = (b-a+1)^{-1}\sum_{j=a}^b \vect{Y}_j$ the mean of the series calculated on the subset implied by  periods $a \leq b$. 
For location $k \in \{2, \dots, p\}$, define the statistic
\begin{equation*}
T_p(k) = \vect{D}_p(k)^\top \mat{V}_p(k)^{-1} \vect{D}_p(k),
\end{equation*}
where
\begin{align*}
\vect{D}_p(k) &= \frac{ (k-1) (p-k+1) }{p^{3/2}}
\left(\widehat{\Btheta}_{1,k-1} - \widehat{\Btheta}_{k,p}\right), \\
\mat{V}_p(k) &= \mat{L}_p(k) + \mat{R}_p(k),
\end{align*}
with
\begin{equation*}
\mat{L}_p(k) =
\left\{
\begin{array}{ll}
\displaystyle \sum_{i=1}^{k-2}
\frac{ i^2 (k-1-i)^2 }{ p^2 (k-1)^2 }
\left( \widehat{\Btheta}_{1,i} - \widehat{\Btheta}_{i+1,k-1} \right)
\left( \widehat{\Btheta}_{1,i} - \widehat{\Btheta}_{i+1,k-1} \right)^\top,
& \qquad \text{if } k > 2, \\
0, & \qquad \text{if } k = 2, \\
\end{array}
\right.
\end{equation*}
and
\begin{equation*}
\mat{R}_p(k) =
\left\{
\begin{array}{ll}
\displaystyle \sum_{i=k+1}^{p}
\frac{ (p-i+1)^2 (i-k)^2 }{ p^2 (k-1)^2 }
\left( \widehat{\Btheta}_{i,p} - \widehat{\Btheta}_{k,i-1} \right)
\left( \widehat{\Btheta}_{i,p} - \widehat{\Btheta}_{k,i-1} \right)^\top,
& \qquad \text{if } k < p, \\
0, & \qquad \text{if } k = p, \\
\end{array}
\right.
\end{equation*}
The final test statistic is then defined as
\begin{equation} \label{eq:SN}
SN_{p} = \max_{k=2,\dots,p} T_p(k).
\end{equation}

Consider a given threshold $K_d > 0$.
If the test statistic exceeds the threshold, $SN_{p} > K_d$, an estimate of the changepoint location is obtained via
\begin{equation*}
\widehat{k} = \argmax_{k=2,\dots,p} T_p(k).
\end{equation*}
Otherwise, no changepoint is detected. 
\citet{shao2010} and \citet{zhao2022} derive theoretical guarantees for this procedure. 
Specifically, \citet{shao2010} obtain the asymptotic distribution of the test statistic $SN_{p}$ for $p \rightarrow \infty$ and provide the corresponding critical values~$K_d$ for a given significance level~$\alpha$ and dimension $d$. 
We summarize relevant critical values in Table~\ref{tab:critvals}.

\begin{table}[!t]
\caption{\textit{Critical values $K_d$ for relevant significance levels~$\alpha$ and dimensions~$d$.}}
\centering
\begin{tabular}{r r r}
\hline\noalign{\smallskip}
 & \multicolumn{2}{c}{Dimension $d$}\\
\noalign{\smallskip}\cline{2-3}\noalign{\smallskip}
$\alpha$  & 1 & 2 \\
\noalign{\smallskip}\hline\noalign{\smallskip}
0.010  & 68.6 & 117.7 \\
0.005 & 84.6 & 135.3 \\
0.001 & 121.9 & 192.5 \\
\noalign{\smallskip}\hline
\end{tabular}
\caption*{\textit{Note.}
The critical values are taken from Table~1 in \citet{shao2010}.}
\label{tab:critvals}
\end{table}

In the context of 
this paper, we use the test statistic from~\eqref{eq:SN} to test for changepoints in the two-dimensional series
\begin{equation*}
\vect{Y}_j =
\begin{pmatrix}
\textsf{\RE}_j\\
\textsf{LSP}_j
\end{pmatrix},
\qquad j = 1,\dots, p,
\end{equation*}
where $\textsf{RE}_j$ is the autoencoder reconstruction error from~\eqref{eq:RE} of a response to item~$j$, and $\textsf{LSP}_j$ is the value of the LSP sequence from~\eqref{eq:LSP} assigned to this response. If a changepoint~$\widehat{k}$ is detected, under Assumptions~\ref{ass:endurance} and~\ref{ass:breaks}, the subsets $\{\vect{Y}_j\}_{j=1}^{\widehat{k}-1}$ and $\{\vect{Y}_j\}_{j=\widehat{k}}^{p}$ may be interpreted as attentive and careless responses, respectively. 
That is, the detected changepoint~$\widehat{k}$ may be viewed as the onset of careless responding. We obtain a series $\{\vect{Y}_j\}_{j=1}^p$ for each of the $n$~survey participants and subsequently test each of the~$n$ series for changepoints.

On a final technical note, LSP sequences may exhibit low variation. 
Computing their variation over a relatively narrow interval---as required by the test statistic $SN_{p}$---may result in a singular (and therefore not invertible) variation matrix $\mat{V}_p(k)$ so that $SN_{p}$ cannot be computed. 
To avoid this issue, we by default inject Gaussian noise of tiny magnitude to each $\textsf{LSP}_j, j=1,\dots,p$. Specifically, we inject draws from a normal distribution with mean zero and variance $0.01^2$. Doing so results in negligible but nonzero variation, which renders~$\mat{V}_p(k)$ invertible.


\clearpage

\section{Additional Simulations} 
\label{app:simulations}

\setcounter{figure}{0}
\setcounter{table}{0}
\setcounter{equation}{0}

\subsection*{Additional Results for the Main Design}

In addition to CODERS, we apply three traditional methods for the detection of careless respondents in the simulation design from the main text, which is inspired by the data collected by \citet{johnson2005}: personal reliability, psychometric antonym, and the longstring index. 
We choose these three methods since they were also applied in the study of \citet{johnson2005}. 
Following \citeauthor{johnson2005}, we reverse psychometric antonym scores,
and we flag respondents as careless whose personal reliability and psychometric antonym scores are lower than $0.3$ and $-0.03$, respectively.
For the longstring index, we flag a respondent as careless if its value exceeds~6 (i.e., $>6$), which is a cutoff value suggested by \citet{niessen2016}.

\begin{figure}[!t]
\caption{\textit{Performance measures for the traditional methods used by \citet{johnson2005}.}}
\centering
\includegraphics[width = \textwidth]{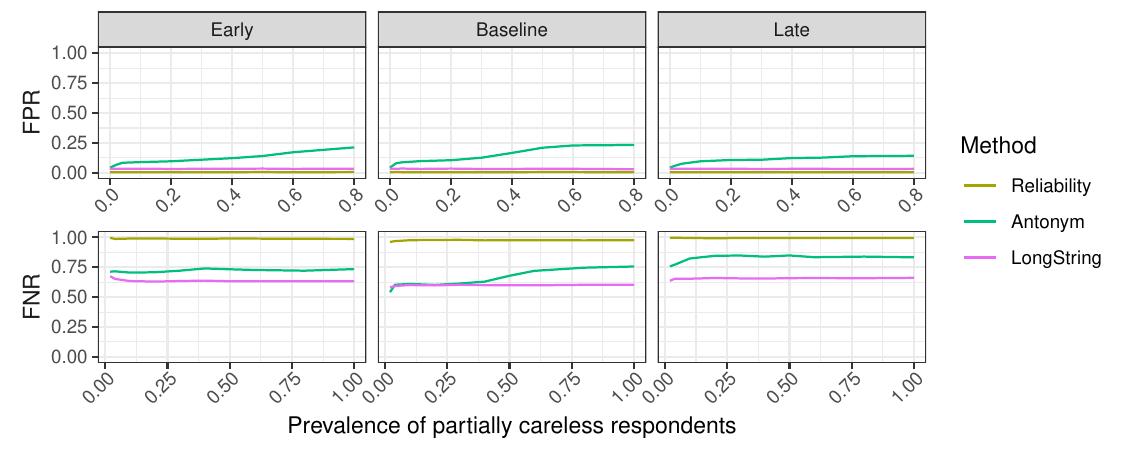}
\caption*{\textit{Note.} The $x$-axis displays various prevalence levels. False positive rates in attentive respondents (top row) and false negative rates in partially careless respondents (bottom row) for different onset regimes (columns) are averaged across 100 repetitions.}
\label{fig:traditional-main}
\end{figure}

\begin{figure}[!t]
\caption{\textit{False negative rate of the traditional methods used by \citet{johnson2005} for different types of partially careless respondents.}}
\centering
\includegraphics[width = \textwidth]{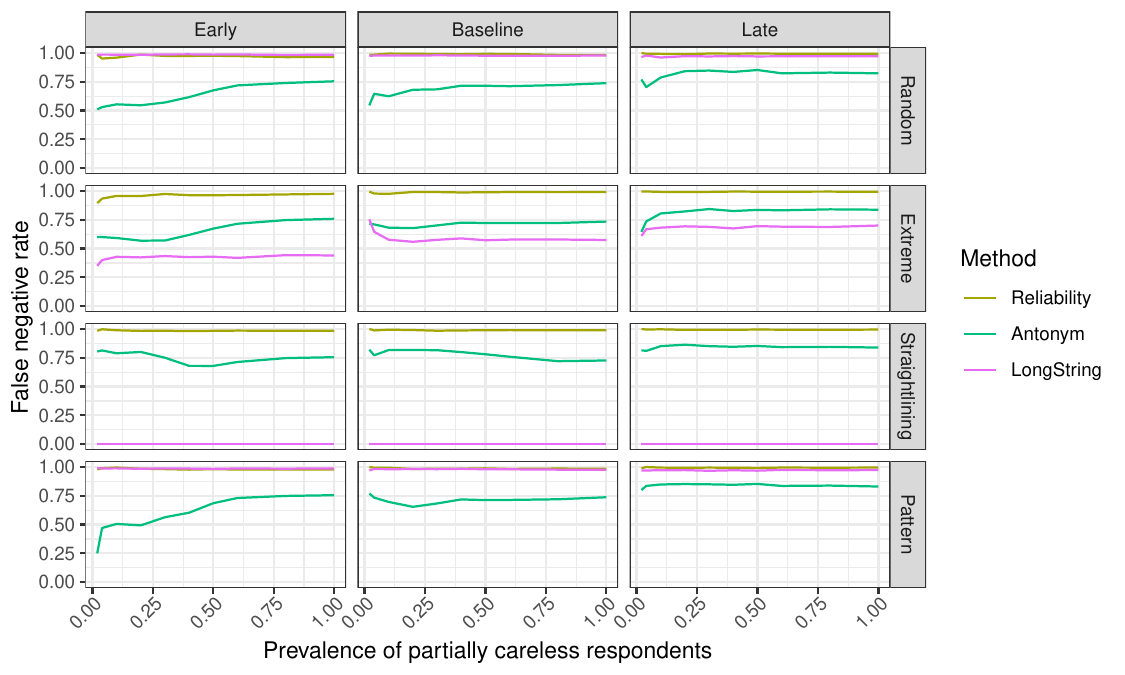}
\caption*{\textit{Note.} The $x$-axis displays various prevalence levels. Results for different types of careless responding (rows) in different onset regimes (columns) are averaged across 100 repetitions.}
\label{fig:traditional-main-subgroups}
\end{figure}

Figure~\ref{fig:traditional-main} visualizes the false positive rate (FPR) and the false negative rate (FNR) of the three methods for the different carelessness onset regimes (in separate columns). 
Figure~\ref{fig:traditional-main-subgroups} further shows the false negative rates for the four types of partially careless respondents (in separate rows).
The longstring index performs perfectly at its intended task of detecting straightlining, whereas it hardly detects any random responding or pattern responding (which is expected). 
Furthermore, it performs better than expected at identifying extreme responding. 
Since the survey is lengthy and those respondents choose randomly from only the two extreme answer categories, there is a relatively high probability (depending on the onset) that they select a long string of the same answer category somewhere by chance. 
It turns out that personal reliability flags hardly any respondents, resulting in false positive rates close to~0 and false negative rates close to~1. 
Psychometric antonym yields inflated false positive rates, while the false negative rates are still relatively high. 
Interestingly, its performance is fairly constant across the different types of carelessness.
As expected, the false negative rate of the traditional methods in general worsens the later the onset of careless responding (while that of CODERS deteriorates the earlier the onset, see Figures~\ref{fig:CODERS-main-careless} and~\ref{fig:CODERS-main-fnr} in the main text).

\clearpage

\subsection*{Decreasing Survey Length}

Recall that the changepoint detection method of \citet{shao2010} used in CODERS is based on asymptotic arguments for the number of items being large (see Appendix~\ref{app:methodology}). 
It therefore seems plausible that the ability of CODERS to accurately identify partial carelessness may decline as survey length decreases.

\subsubsection*{Data Generation}

We use the same simulation design of the main text, but we focus on the baseline onset regime and instead decrease the number of items $p$ by gradually removing sets of items. 
Specifically, for a given simulated dataset, we remove all items belonging to randomly selected personality traits. 
There is a total of five traits, each of which is measured by six facets comprising ten items each (300 items in total). 
Hence, removing a single trait from the measurement will remove $6 \times 10=60$ items. 
We drop up to four traits, resulting in surveys comprising a total of 
$p \in \{ 240, 180, 120, 60 \}$ items.

\subsubsection*{Results}

\begin{figure}[!t]
 \caption{\textit{False positive rate of CODERS for attentive respondents across decreasing survey length.}}
 \centering
  \includegraphics[width = \textwidth]{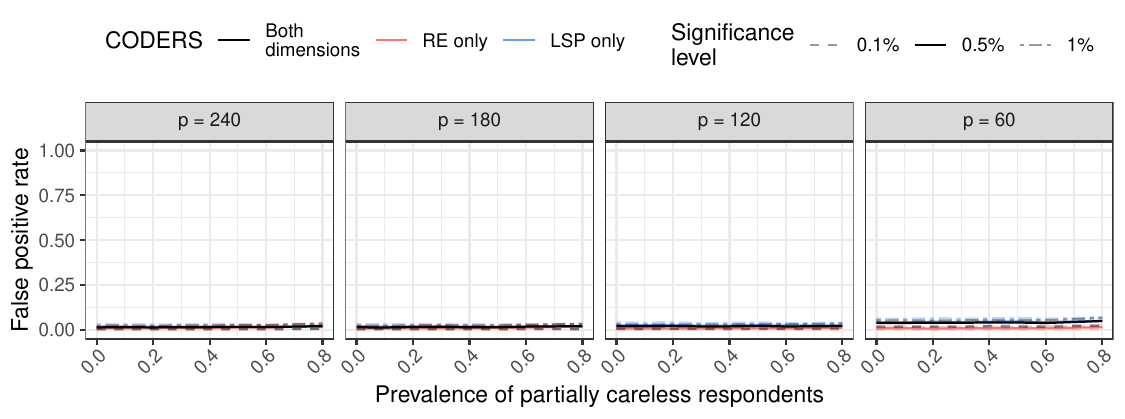}
  \caption*{\textit{Note.} The $x$-axis displays various prevalence levels of partially careless respondents. Results for different survey lengths are shown in separate columns, averaged across 100 repetitions.}
  \label{fig:CODERS-shortsurveys-fpr}
\end{figure}

\begin{figure}[!b]
 \caption{\textit{Performance measures of CODERS for partially careless respondents across decreasing survey length.}}
 \centering
  \includegraphics[width = \textwidth]{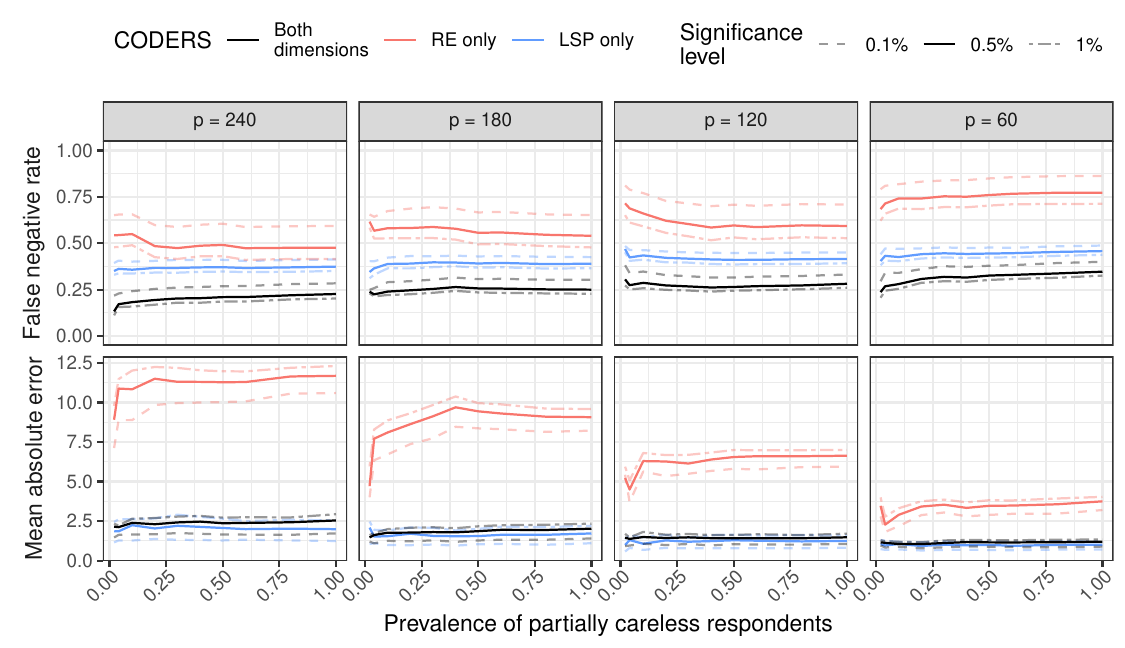}
  \caption*{\textit{Note.} The $x$-axis displays various prevalence levels. The top row shows results for the false negative rate, while the bottom row visualizes the accuracy of onset estimation via the mean absolute error. Results for different survey lengths are shown in separate columns, averaged across 100 repetitions.}
  \label{fig:CODERS-shortsurveys-careless}
\end{figure}

Figure \ref{fig:CODERS-shortsurveys-fpr} plots the false positive rates for the different survey lengths (in separate columns). 
As before, CODERS rarely incorrectly flags attentive respondents as careless at any of the considered significance levels, although there is some overrejection for the shortest survey with $p = 60$. 
Figure~\ref{fig:CODERS-shortsurveys-careless} visualizes the false negative rate (top row) and the mean absolute error of onset estimation (bottom row) for careless respondents across the different surveys lengths (in separate columns). 
At a given significance level, the false negative rate slightly increases with decreasing survey length compared to the results from the main text. 
For $p = 60$, the false negative rate reaches about 30\%--40\%, depending on the significance level. 
In addition, the location of the onset of carelessness is still estimated highly accurately throughout the investigated survey lengths, on average being off by only about one or two items.

\begin{figure}[!b]
\caption{\textit{False negative rate of CODERS for different types of partially careless respondents across decreasing survey length.}}
\centering
\includegraphics[width = \textwidth]{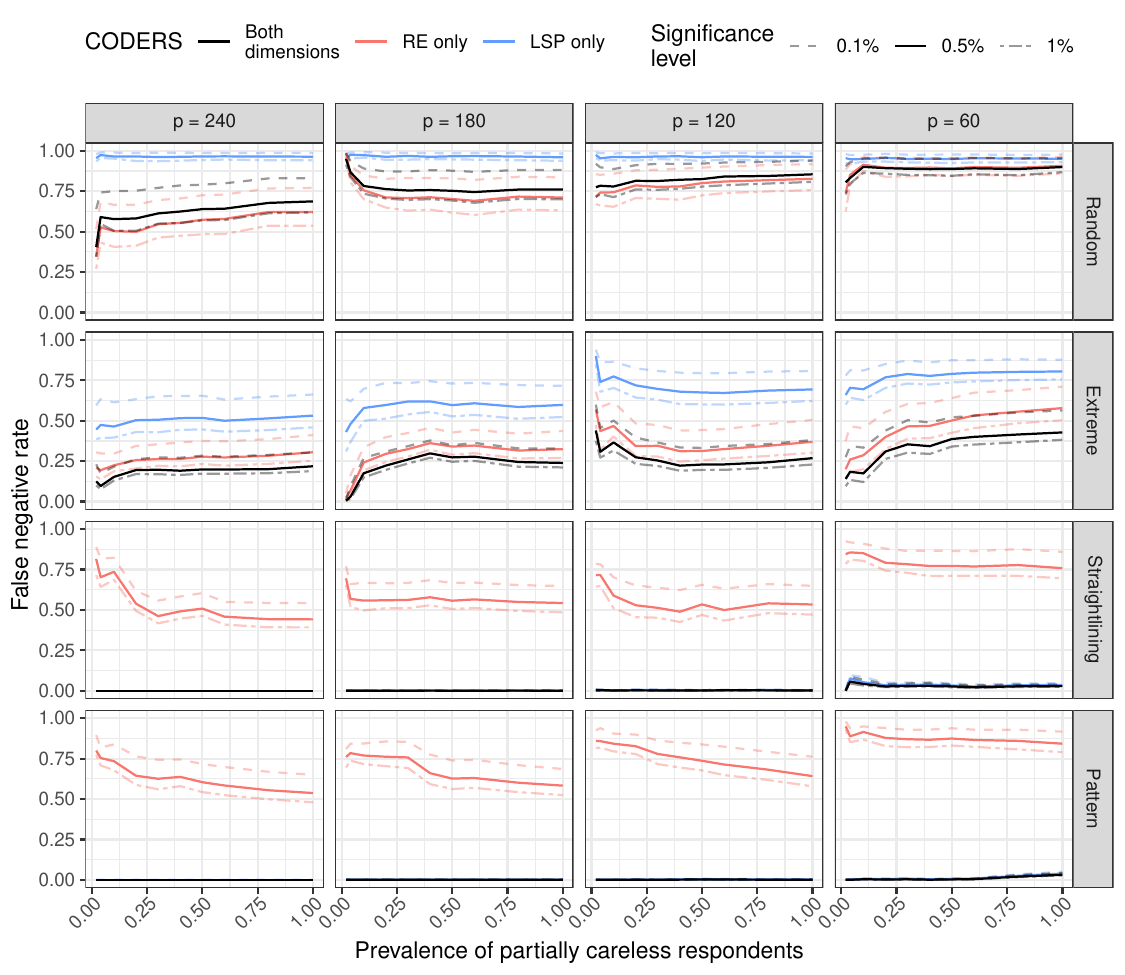}
\caption*{\textit{Note.} The $x$-axis displays various prevalence levels. Results for different types of careless responding (rows) in different survey lengths (columns) are averaged across 100 repetitions.}
\label{fig:CODERS-shortsurveys-fnr}
\end{figure}

\begin{figure}[!t]
\caption{\textit{Mean absolute error of carelessness onset estimation with CODERS for different types of partially careless respondents across decreasing survey length.}}
\centering
\includegraphics[width = \textwidth]{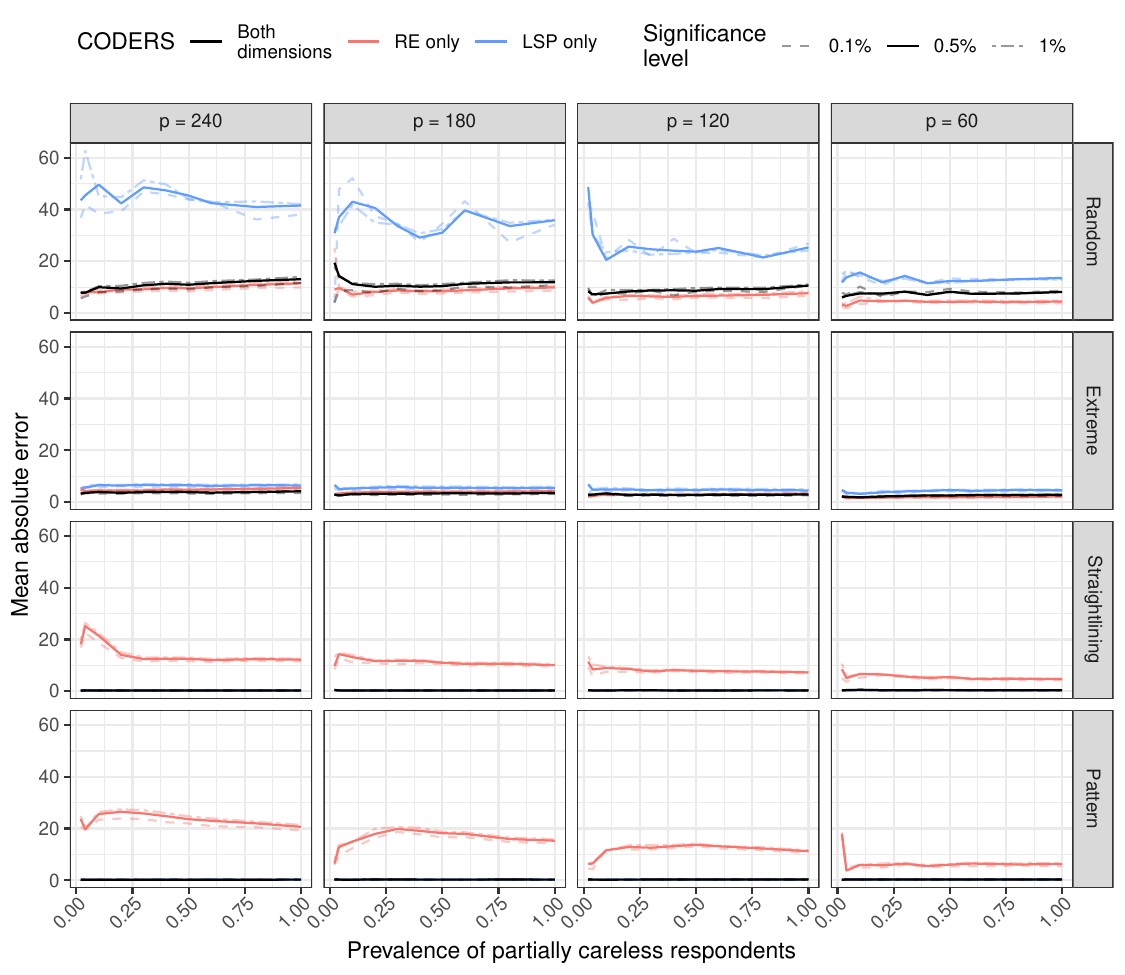}
\caption*{\textit{Note.} The $x$-axis displays various prevalence levels. Results for different types of careless responding (rows) in different survey lengths (columns) are averaged across 100 repetitions.}
\label{fig:CODERS-shortsurveys-mae}
\end{figure}

Figures~\ref{fig:CODERS-shortsurveys-fnr} and~\ref{fig:CODERS-shortsurveys-mae} show the false negative rate and the mean absolute error of onset estimation, respectively, grouped by carelessness type (rows) for each considered survey length (columns). 
In essence, the slight increase in false negative rates seems to be primarily caused by fewer inconsistent careless respondents being detected, whereas the performance for invariable careless responding stays near-perfect.
For $p = 60$, in particular very few random respondents are correctly identified, but the performance is still acceptable for extreme respondents.

We conclude that, as expected, the power of CODERS for detecting partially careless respondents drops as the survey length decreases. 
Yet, the drop in power is small and primarily caused by a low detection performance for random respondents. 
It is conceivable that reconstruction errors are a noisier measurement of inconsistency (compared to LSP as a measurement of invariability), as the autoencoder has less information to learn attentive responding behavior in shorter surveys, explaining why the loss of power is primarily visible for random responding. 
Nevertheless, if a changepoint is flagged, its accuracy is nearly unchanged at excellent levels. 
That is, even though CODERS has somewhat less power in identifying partially careless respondents, the onset of carelessness in the flagged respondents is sill accurately estimated. 
Furthermore, the power loss can be partially alleviated by choosing a less conservative significance level, such as~0.5\% or~1\%.

\clearpage

\subsection*{Additional Sample Sizes} 

Although the number of items in the survey may be the primary driver regarding power of the changepoint detection test used in CODERS, the autoencoder may also benefit from more information in the form of a larger sample size in order to learn attentive response behavior (and conversely suffer under a smaller sample size).

\subsubsection*{Data Generation}

We again use the same simulation design of the main text and focus on the baseline onset regime, but we now generate a smaller and a larger number of participants, namely $n=248$ and $n=\text{1,000}$, respectively. 
Note that the sample size in our simulations should be divisible by~4 since we generate four types of partially careless respondents. This explains the perhaps unusual choice of $n = 248$ (rather than $n = 250$).

\subsubsection*{Results}

\begin{figure}[!t]
 \caption{\textit{False positive rate of CODERS for attentive respondents across additional sample sizes.}}
 \centering
  \includegraphics[width = \textwidth]{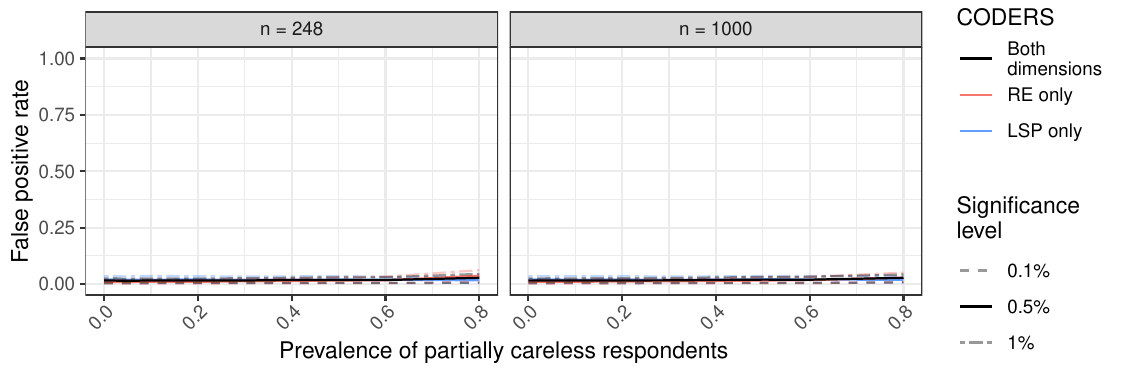}
  \caption*{\textit{Note.} The $x$-axis displays various prevalence levels of partially careless respondents. Results for different sample sizes are shown in separate columns, averaged across 100 repetitions.}
  \label{fig:CODERS-varyn-fpr}
\end{figure}

\begin{figure}[!b]
 \caption{\textit{Performance measures of CODERS for partially careless respondents across additional sample sizes.}}
 \centering
  \includegraphics[width = \textwidth]{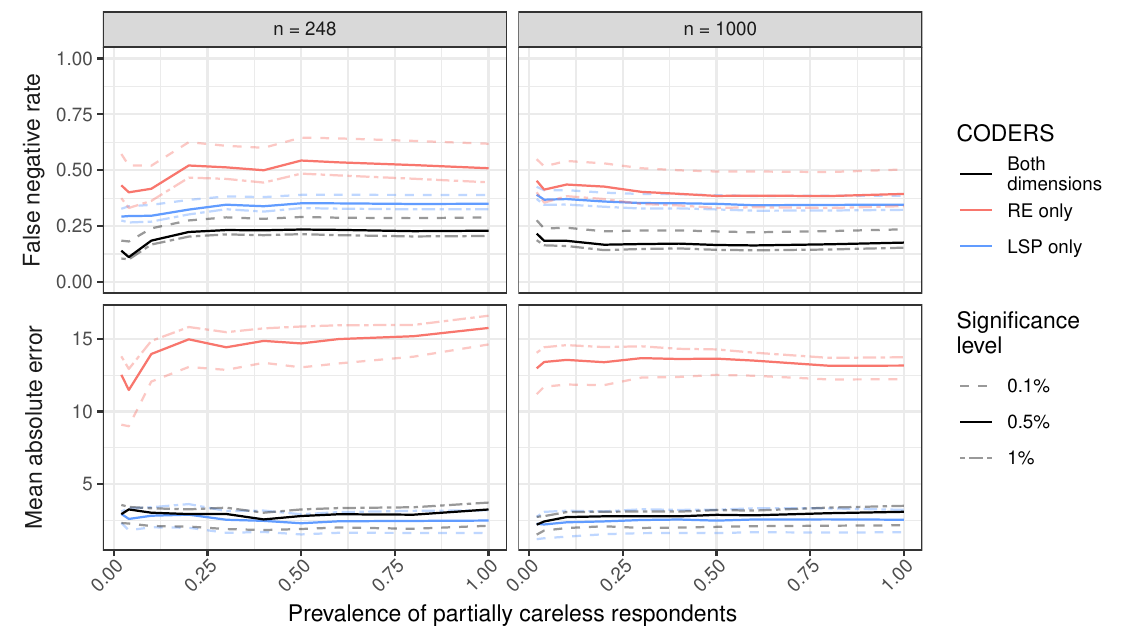}
  \caption*{\textit{Note.} The $x$-axis displays various prevalence levels. The top row shows results for the false negative rate, while the bottom row visualizes the accuracy of onset estimation via the mean absolute error. Results for different sample sizes are shown in separate columns, averaged across 100 repetitions.}
  \label{fig:CODERS-varyn-careless}
\end{figure}

\begin{figure}[!t]
\caption{\textit{False negative rate of CODERS for different types of partially careless respondents across additional sample sizes.}}
\centering
\includegraphics[width = \textwidth]{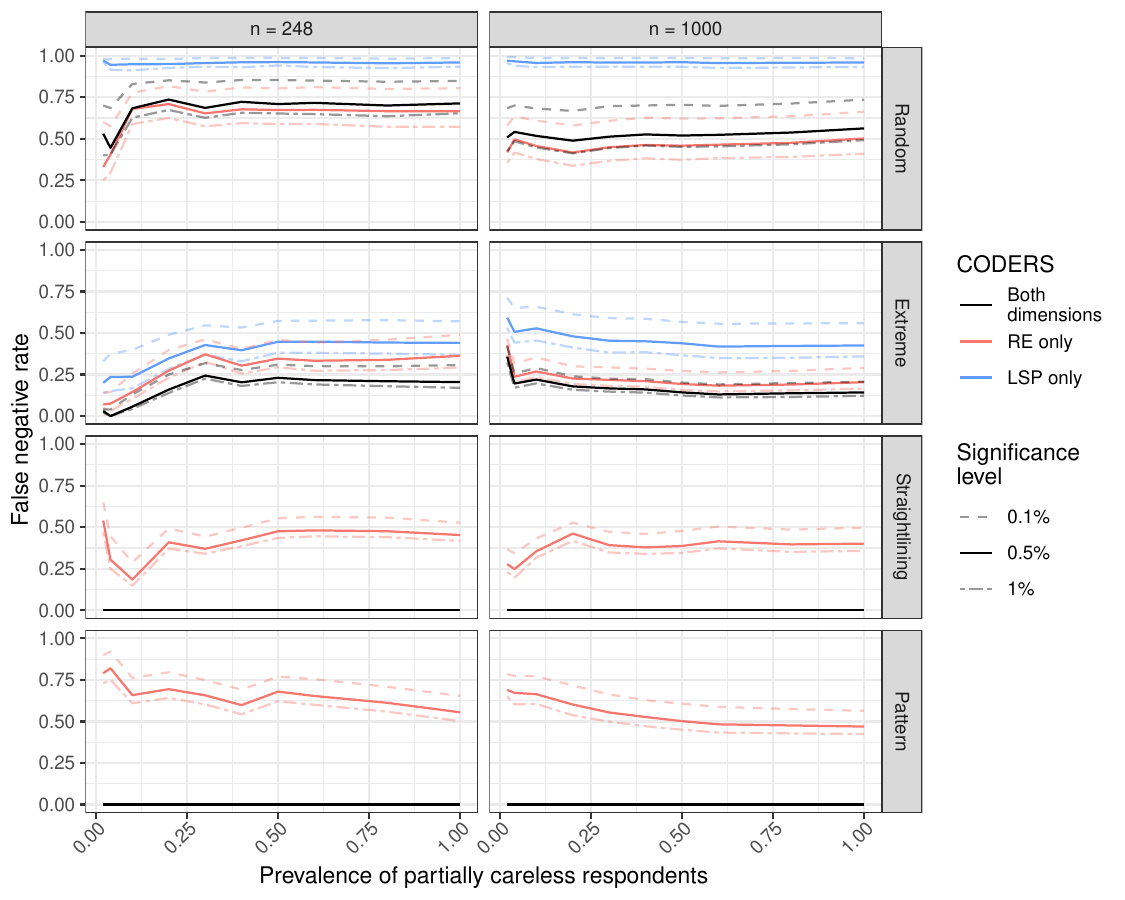}
\caption*{\textit{Note.} The $x$-axis displays various prevalence levels. Results for different types of careless responding (rows) in different sample sizes (columns) are averaged across 100 repetitions.}
\label{fig:CODERS-varyn-fnr}
\end{figure}

\begin{figure}[!t]
\caption{\textit{Mean absolute error of carelessness onset estimation with CODERS for different types of partially careless respondents across additional sample sizes.}}
\centering
\includegraphics[width = \textwidth]{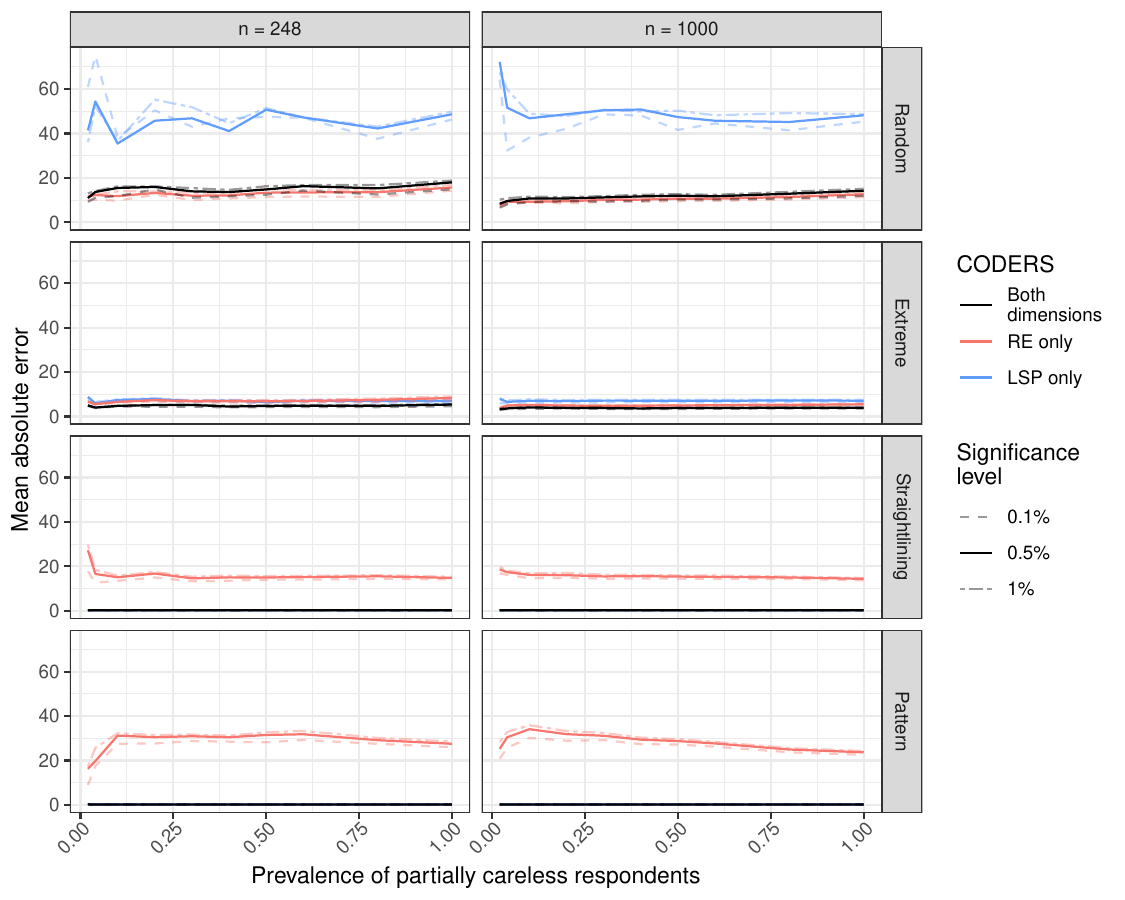}
\caption*{\textit{Note.} The $x$-axis displays various prevalence levels. Results for different types of careless responding (rows) in different sample sizes (columns) are averaged across 100 repetitions.}
\label{fig:CODERS-varyn-mae}
\end{figure}

Figures~\ref{fig:CODERS-varyn-fpr}--\ref{fig:CODERS-varyn-mae} visualize the performance measures for $n=248$ and $n=1,000$ simulated participants (in separate columns). 
The results are similar to those in the main text, with the overall FNR in partially careless respondents increasing only slightly for the smaller sample size and decreasing only slightly for the larger sample size. For random respondents, however, the increase in false negatives in the smaller sample size is notable, but so is the decrease in FNR in the larger sample size (see Figure~\ref{fig:CODERS-varyn-fnr}, top row). 
This can likely be explained by the amount of information the autoencoder has available in order to learn to distinguish between attentive responding and random responding. 
It is therefore conceivable that the power for detecting random respondents may improve further in even larger sample sizes.

\clearpage

\subsection*{Randomly Choosing from the Middle Answer Categories}

Recall that the simulation design in the main text considers four types of careless responding: random responding, extreme responding, pattern responding, and straightlining. 
However, there may be other relevant types of carelessness beyond those four, including ones that are extremely difficult to distinguish from attentive response behavior.

\subsubsection*{Data Generation}

As before, we take the simulation design from the main text with the baseline onset regime, but instead of the four aforementioned types of careless responding, we generate a distinct type that we refer to as \emph{middling}. 
Middling is characterized by randomly choosing between the three middle answer categories (out of five).
As such, middling carelessness can be seen as an inconsistent response type complementing extreme responding.
The major difference to extreme carelessness is that central response options are more frequently chosen by attentive respondents in our simulation design. 
In fact, with the chosen marginal distributions and correlation structure (see Figure~\ref{fig:cormat-marginals}), there is a lot of natural variation among the central response options in attentive responses. 
Hence, it is difficult to systematically distinguish middling carelessness from attentive responding, but there is also the risk that attentive respondents might be mistaken for middling careless respondents.

\subsubsection*{Results}

\begin{figure}[!t]
 \caption{\textit{False positive rate of CODERS for attentive respondents in the presence of partially careless respondents that randomly choose from the middle answer categories.}}
 \centering
  \includegraphics[width = \textwidth]{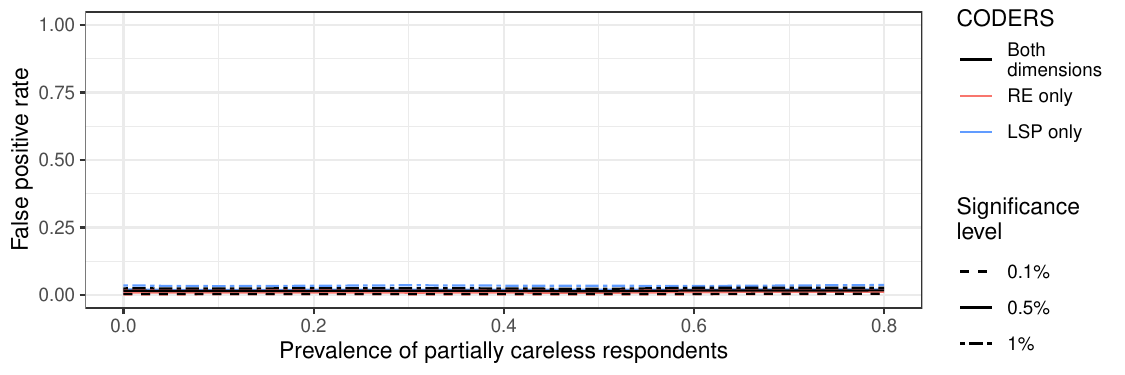}
  \caption*{\textit{Note.} The $x$-axis displays various prevalence levels of partially careless respondents. Results are averaged across 100 repetitions.}
  \label{fig:CODERS-middling-fpr}
\end{figure}

\begin{figure}[!b]
 \caption{\textit{Performance measures of CODERS for partially careless respondents that randomly choose from the middle answer categories.}}
 \centering
  \includegraphics[width = \textwidth]{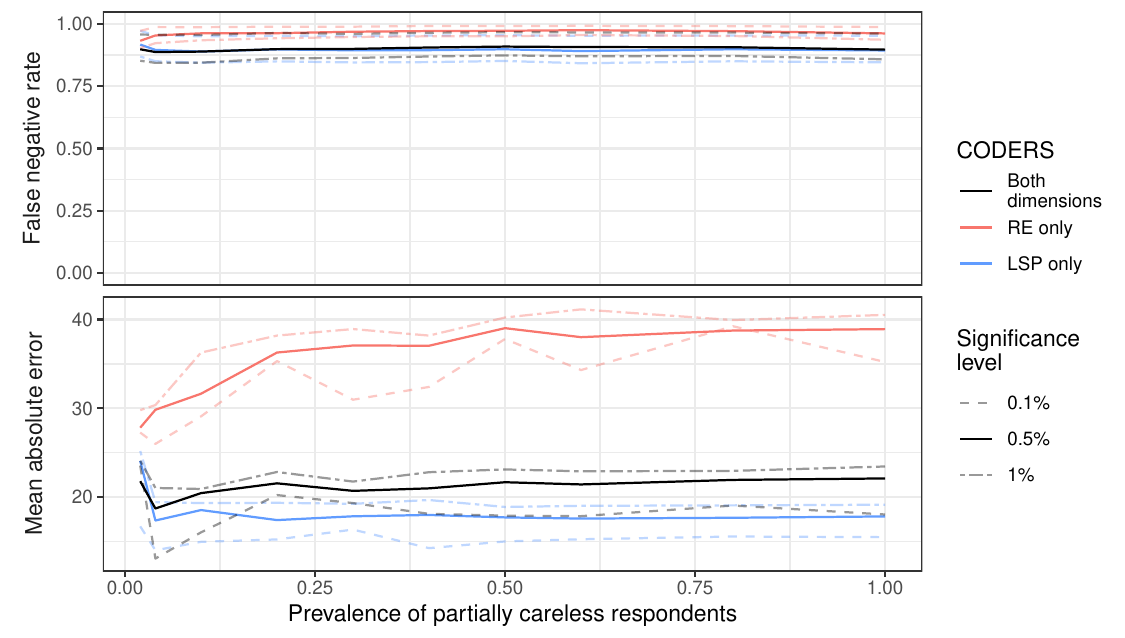}
  \caption*{\textit{Note.} The $x$-axis displays various prevalence levels. The top row shows results for the false negative rate, while the bottom row visualizes the accuracy of onset estimation via the mean absolute error. Results are averaged across 100 repetitions.}
  \label{fig:CODERS-middling-careless}
\end{figure}

Figures~\ref{fig:CODERS-middling-fpr} and~\ref{fig:CODERS-middling-careless} visualizes the performance of CODERS when partially careless respondents are of the middling type. 
One the one hand, CODERS still almost never incorrectly flags an attentive respondent as partially careless. 
On the other hand, it identifies very few middling respondents with a false negative rate between about 85\% and 95\%, depending on the significance level. 
Hence, these simulations further demonstrate that CODERS requires inconsistent responding to be sufficiently different from attentive responding in order to distinguish between the two.

\begin{figure}[!t]
\caption{\textit{Performance measures of the traditional methods for partially careless respondents that randomly choose from the middle answer categories.}}
\centering
\includegraphics[width = \textwidth]{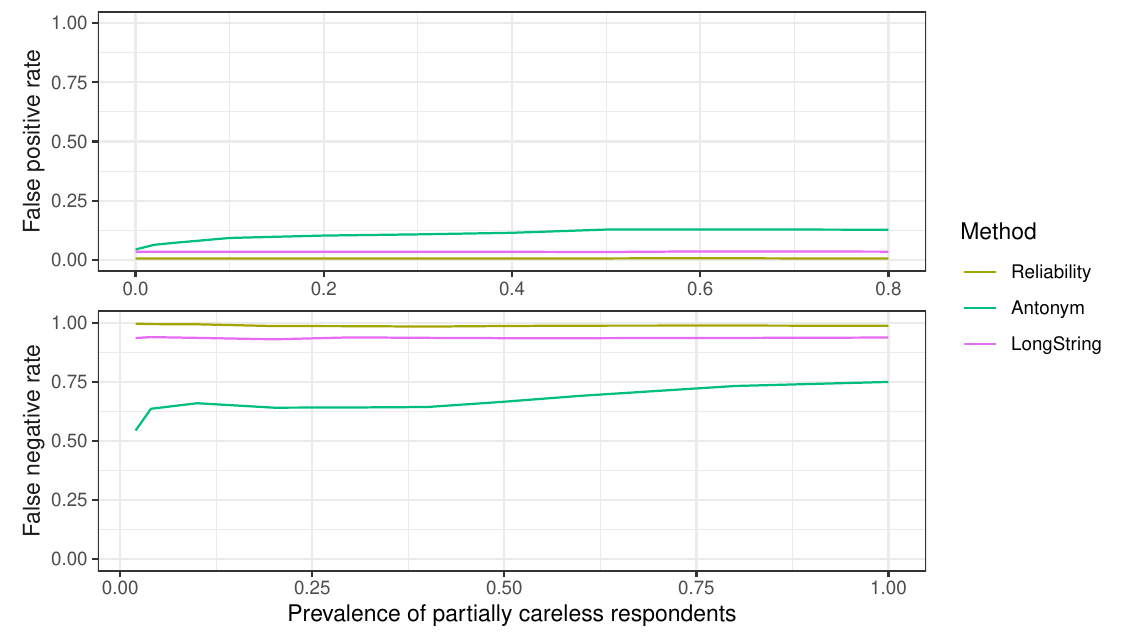}
\caption*{\textit{Note.} The $x$-axis displays various prevalence levels. False positive rates in attentive respondents (top row) and false negative rates in partially careless respondents (bottom row) are averaged across 100 repetitions.}
\label{fig:traditional-middling}
\end{figure}

In order to further highlight the difficulty of detecting middling carelessness, we again apply the traditional methods personal reliability, psychometric antonym, and the longstring index. 
The false positive rates and false negative rates are shown in Figure~\ref{fig:traditional-middling}.
As in the simulation design from the main text, personal reliability flags hardly any respondents, resulting in false positive rates close to~0 and false negative rates close to~1. The performance of the longstring index is similar, which is not surprising because it was designed for detecting straightlining (invariable carelessness) rather than inconsistent responding like middling. 
Psychometric antonym identifies more middling respondents than CODERS with a false negative rate of about 65\%--75\%, but it exhibits an inflated false positive rate of up to about~15\%.

\clearpage

\subsection*{Temporary Carelessness}

It seems plausible that more than one changepoint may occur in a participant's response behavior. 
For instance, a respondent may start the survey attentively, but become careless at some point due to being distracted, only to start responding attentively again once the distraction vanishes.

\subsubsection*{Data Generation}

We use the same simulation design from the main text, but instead of simulating only one changepoint, we simulate temporary carelessness as follows. 
We randomly sample the first changepoint from the item set $\{30,31,\dots,149,150\}$ (between~10\% and~50\% of all items) and the second changepoint from the item set $\{151,152,\dots, 269,270\}$ (between~50\% and~90\% of all items). Between these two changepoints, we generate careless responses using the same four types of carelessness from the main text.

\subsubsection*{Results}

\begin{figure}[!t]
 \caption{\textit{False positive rate of CODERS for attentive respondents in case of temporary carelessness.}}
 \centering
  \includegraphics[width = \textwidth]{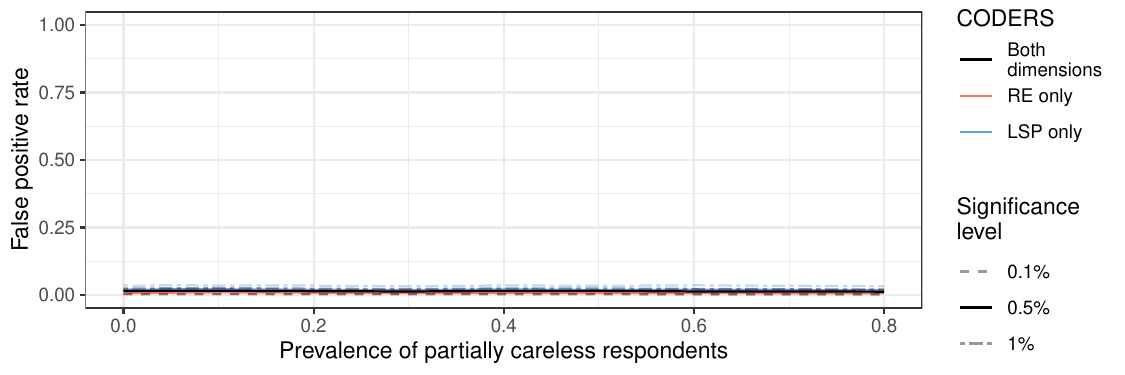}
  \caption*{\textit{Note.} The $x$-axis displays various prevalence levels of partially careless respondents. Results are averaged across 100 repetitions.}
  \label{fig:CODERS-temporary-fpr}
\end{figure}

\begin{figure}[!b]
 \caption{\textit{Performance measures of CODERS for partially careless respondents in case of temporary carelessness.}}
 \centering
  \includegraphics[width = \textwidth]{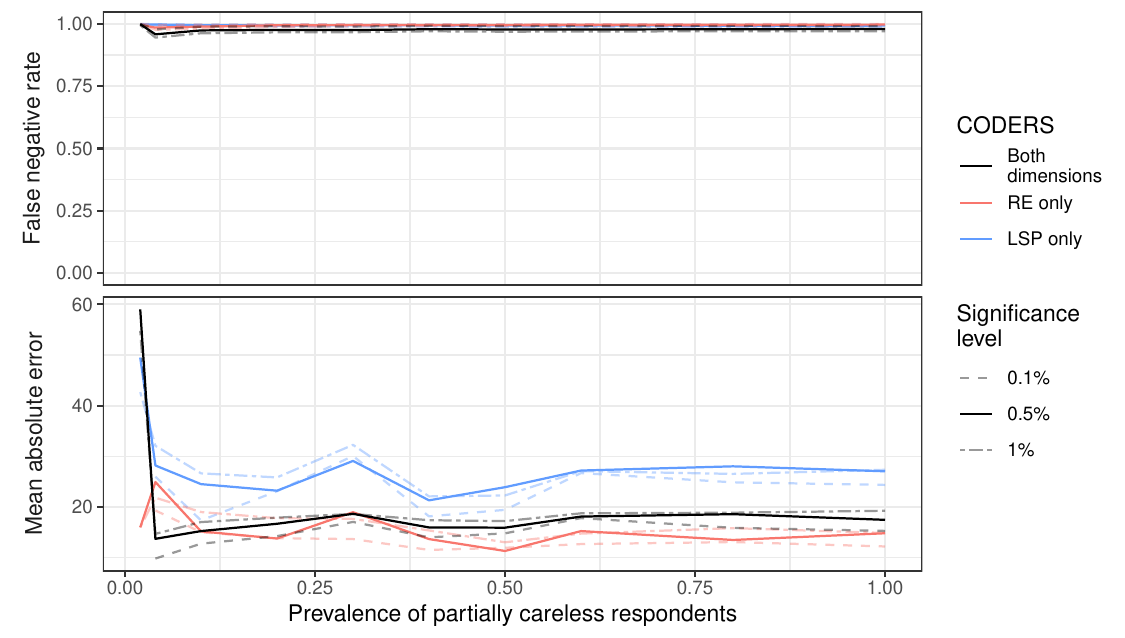}
  \caption*{\textit{Note.} The $x$-axis displays various prevalence levels. The top row shows results for the false negative rate, while the bottom row visualizes the accuracy of onset estimation via the mean absolute error. Results are averaged across 100 repetitions.}
  \label{fig:CODERS-temporary-careless}
\end{figure}

\begin{figure}[!t]
\caption{\textit{Performance measures for the traditional methods in case of temporary carelessness.}}
\centering
\includegraphics[width = \textwidth]{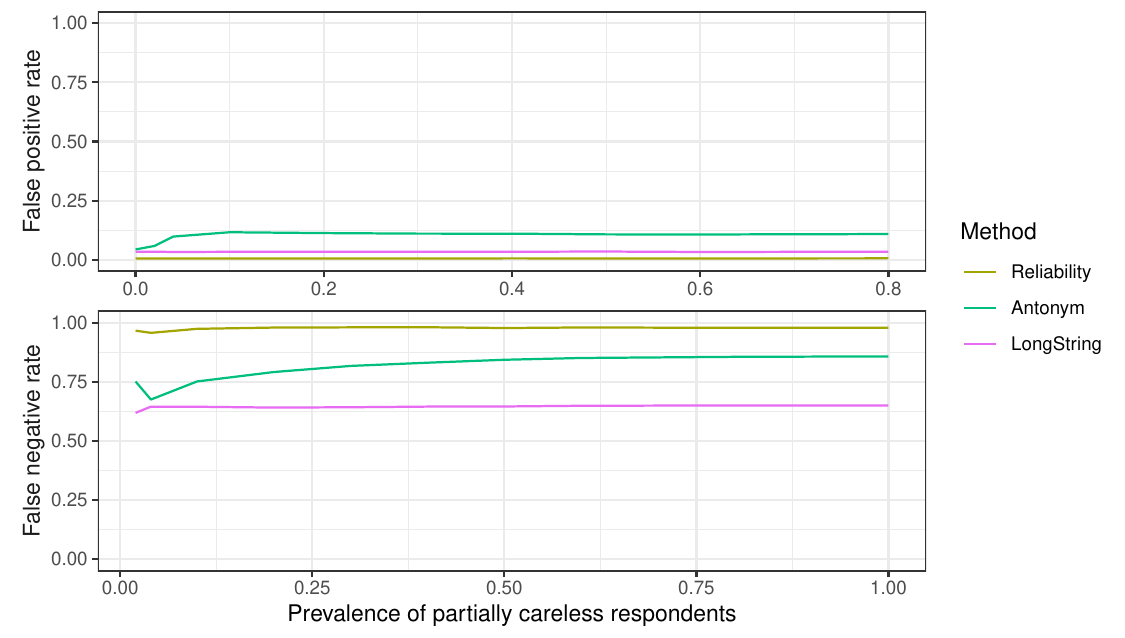}
\caption*{\textit{Note.} The $x$-axis displays various prevalence levels. False positive rates in attentive respondents (top row) and false negative rates in partially careless respondents (bottom row) are averaged across 100 repetitions.}
\label{fig:traditional-temporary}
\end{figure}

\begin{figure}[!b]
\caption{\textit{False negative rate of the traditional methods for different types of partially careless respondents in case of temporary carelessness.}}
\centering
\includegraphics[width = \textwidth]{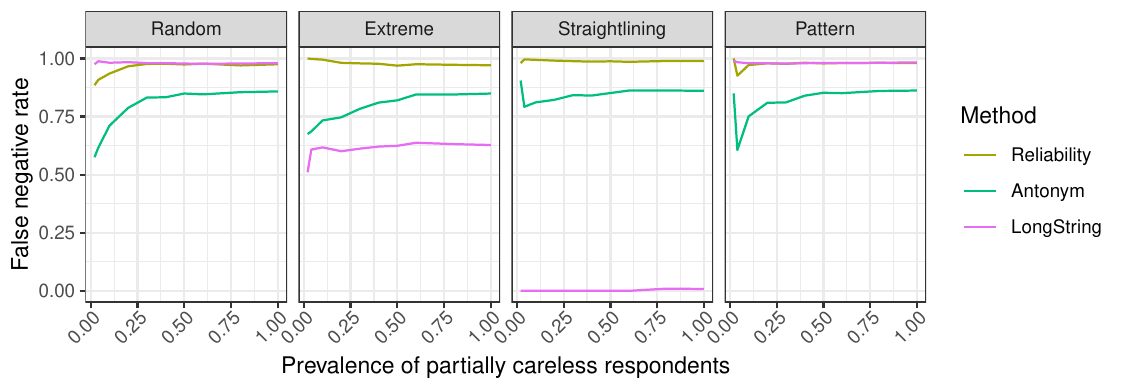}
\caption*{\textit{Note.} The $x$-axis displays various prevalence levels. Results for different types of careless responding are shown in separate columns, averaged across 100 repetitions.}
\label{fig:traditional-temporary-subgroups}
\end{figure}

Figures~\ref{fig:CODERS-temporary-fpr} and~\ref{fig:CODERS-temporary-careless} visualize the performance of CODERS and Figures~\ref{fig:traditional-temporary} and~\ref{fig:traditional-temporary-subgroups} that of the traditional methods. 
Clearly, CODERS rarely flags any participants, with a false positive rate close to~0 and a false negative rate close to~1 (we therefore omit separate results for the four types of partially careless responding). 
The same behavior is observed for personal reliability. 
Psychometric antonym achieves a better false negative rate (about 70\%--85\%) but struggles with many false positives (up to about 15\%). 
The longstring index yields somewhat better performance, with a relatively low false positive rate and a false negative rate around 65\%. 
As in the simulation design from the main text, it successfully captures straightliners as well as some extreme respondents (see the corresponding section above for a more detailed discussion). 
In summary, CODERS is simply not designed for detecting more than one changepoint in respondent behavior, such that an extension to multiple changepoint detection may be a fruitful direction for further research.

\clearpage

\section{Details on the Empirical Application} 
\label{app:application}

\setcounter{figure}{0}
\setcounter{table}{0}
\setcounter{equation}{0}

\subsection*{Data Preprocessing}

Among other information, the data of \citet{johnson2005}, provided at \url{https://osf.io/sxeq5/}, contain per-item responses to~$p=300$ IPIP-NEO-300 items for each of the the~$n=\text{20,993}$ respondents.
The responses in the provided data have already underwent some preprocessing that is described in detail the accompanying README file. 
Specifically, responses to negatively-worded items were reverse-coded, that is, a given response of~5 was recoded to~1, 4 was recoded to~3, and so on. 
Thus, in order to recover the original responses, we reverted this recoding. 
Furthermore, a total of~33,691 individual responses are missing (0.53\% of all responses). 
Note that \citeauthor{johnson2005} had already removed participants with too many missing responses from the provided data set, hence there are relatively few missing responses for the remaining participants.
Due to the low number of missing responses, we treat them as a separate answer category by coding them as~0 when applying CODERS and the traditional screeners for careless responding.

\subsection*{Additional Results}

\begin{figure}[!t]
 \caption{\textit{Histogram of average reconstruction errors for the data of \citet{johnson2005}.}}
 \centering
  \includegraphics[width = \textwidth]{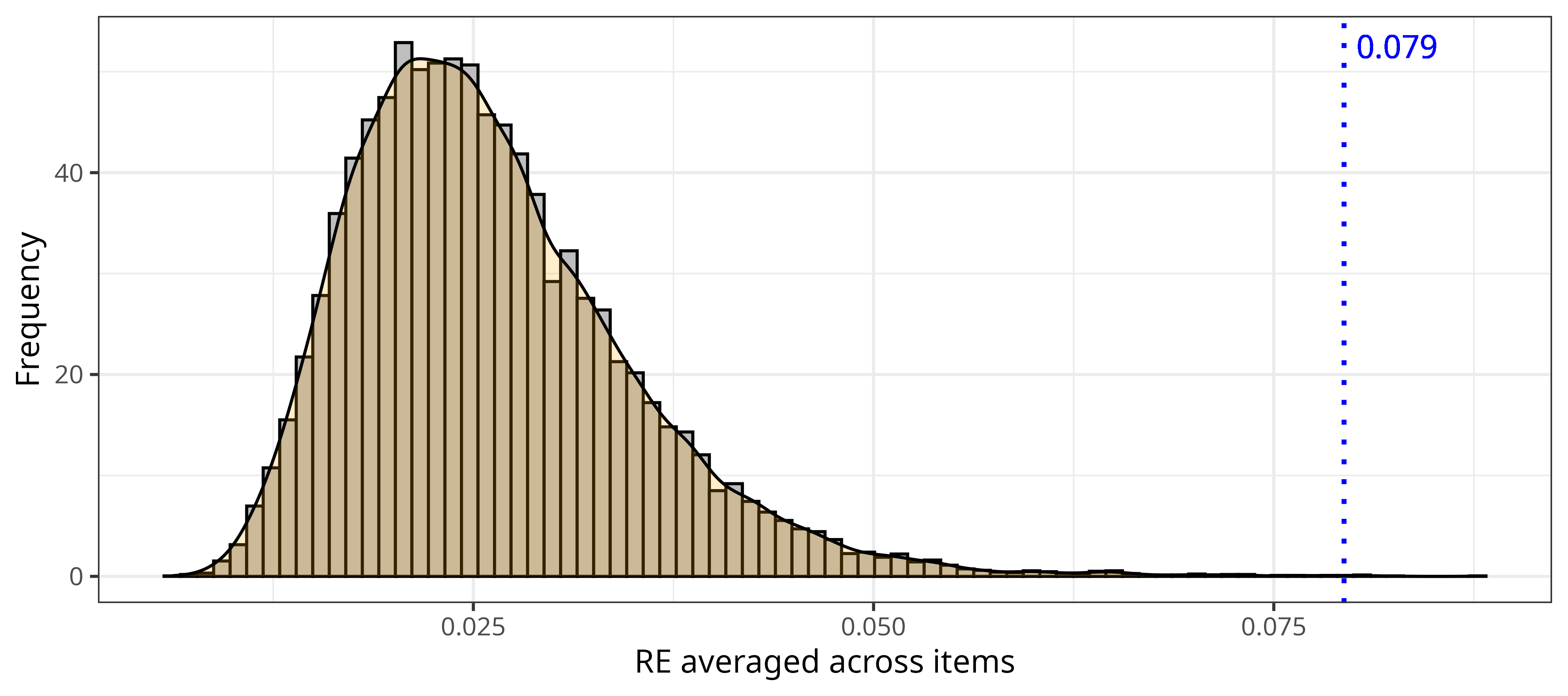}
  \caption*{\textit{Note.} 
  The $x$-axis contains the per-respondent average reconstruction errors over the~$p=300$ items. On the $y$-axis, a histogram of the average reconstruction errors for $n=\text{20,993}$ respondents is displayed together with a density estimate. The vertical dotted line denotes the average reconstruction error for the participant in Figure~\ref{fig:appl-inconsistent-throughout} and Table~\ref{tab:appl-inconsistent-throughout} in the main text. This value, 0.079, lies in the~99.98-th percentile of all average reconstruction errors.}
 \label{fig:johnson2005density}
 \end{figure}

Figure~\ref{fig:johnson2005density} provides a histogram as well as a density estimate for the per-respondent average reconstruction errors $\textsf{RE}_{i} = \frac{1}{p} \sum_{j=1}^p \textsf{RE}_{ij}$, $i=1,\dots, n$, where averages are taken over the~$p=300$ items. We can see that the density has a long right tail, which might be due to careless respondents. We highlight the value 0.079 by a vertical dotted line, which is the average reconstruction error of the participant from Figure~\ref{fig:appl-inconsistent-throughout} and Table~\ref{tab:appl-inconsistent-throughout} in the main text. This value is in the~99.98-th percentile of all~$n=\text{20,993}$ average reconstruction errors, with only 6~participants having a higher value. Given the extremely high average reconstruction error and the fact that this respondent is also flagged by psychometric antonym, it is likely that the participant responded carelessly throughout the survey, explaining why CODERS did not flag a changepoint.
 
\end{document}